\documentclass{aa}  
\usepackage{amsmath}
\usepackage{comment}
\usepackage{graphicx}
\usepackage{txfonts}
\usepackage{color}
\usepackage{xcolor}
\usepackage[colorlinks]{hyperref}

\hypersetup{
    colorlinks = true,
    linkcolor = {blue},
    citecolor = {blue},
}

\usepackage{array}
\usepackage{multirow}
\usepackage{natbib}
\newcolumntype{P}[1]{>{\centering\arraybackslash}p{#1}}

\usepackage{enumitem}
\usepackage{amsmath}

\begin{document}

   \title{Far-ultraviolet flux distribution in Orion and its relation to stellar accretion}


   \author{Rossella Anania
          \inst{1,4} \thanks{email: ananiar@tcd.ie},
          Andrew J. Winter
          \inst{2},
          Miguel Vioque
          \inst{3},
          Giovanni P. Rosotti
          \inst{4},
          Giacomo Beccari
          \inst{3},
          Giuseppe Lodato
          \inst{4},
          Lorenzo A. Malanga
          \inst{4},
          Lara Piscarreta
          \inst{3},
          Alice Somigliana
          \inst{5},
          Leonardo Testi
          \inst{6},
          Claudia Toci
          \inst{7}
          }

   \institute{School of Physics, Trinity College Dublin, the University of Dublin, College Green, Dublin 2, Ireland
        \and
            Astronomy Unit, School of Physics and Astronomy, Queen Mary University of London, London E1 4NS, UK
        \and
            European Southern Observatory, Karl-Schwarzschild-Str. 2, 85748 Garching bei München, Germany
        \and
            Dipartimento di Fisica, Università degli Studi di Milano, Via Celoria 16, I-20133 Milano, Italy
        \and
            MaxPlanck Institute for Astronomy, Königstuhl 17, 69117 Heidelberg, Germany
        \and
            Alma Mater Studiorum– Università di Bologna, Dipartimento di Fisica e Astronomia “Augusto Righi”, Via Gobetti 93/2, 40129 Bologna, Italy
        \and
            Escuela Técnica Superior de Ingeniería, Universidad de Sevilla, Camino de los Descubrimientos s/n, 41092 Sevilla, Spain 
            }

   \date{(Accepted June 26, 2026)}

 
  \abstract{
  \textit{Context.} Orion is the closest region hosting active star formation and young OBA stars. Accurately determining the far-ultraviolet (FUV) flux at its stellar population is essential to connect stellar and protoplanetary disc properties to the environment.\\
  \textit{Aims.} We (1) accurately estimated the FUV flux and its distribution at a numerous stellar population of Orion by statistically accounting for the uncertainty in parallax measurements, and (2) investigated the relation between stellar accretion and external FUV radiation field by comparing observations and disc evolution models.\\
  \textit{Methods.} 
  We selected a large stellar population in Orion (within a $6^\circ$ radius of the Orion Nebula Cluster core), assigned sub-cluster memberships, and 
  used the two-dimensional sub-cluster geometry to infer three-dimensional separations from OBA stars and compute the FUV flux (and its uncertainty) at each stellar position. 
  We studied the accretion luminosities ($L_{\mathrm{acc}}$) inferred from H$_{\alpha}$ emission in Gaia XP spectra of Orion sources and determined their detection fraction as a function of age and FUV flux. We compared the results with population synthesis models of viscous discs experiencing external photoevaporation.\\
  \textit{Results.} We provided a publicly available table of FUV fluxes at $\sim 8600$ stars in Orion. Most of this stellar population is weakly FUV-irradiated, $<10^{2} \ \mathrm{G}_{0}$, $\sim35\%$  is intermediately irradiated, $10^{2}-10^{4} \ \mathrm{G}_{0}$, and only $\sim 5\%$ has FUV fluxes $> 10^{4} \ \mathrm{G}_{0}$. 
  Gaia-based $L_{\mathrm{acc}}$ decreases with age, and H$_{\alpha}$ detection fraction declines more rapidly in regions with strong FUV fluxes ($\gtrsim 10^{2} \ \mathrm{G}_{0}$) than in regions exposed to weaker FUV fluxes ($\lesssim10^{2} \ \mathrm{G}_{0}$), broadly consistent with the model. This result may suggest that external photoevaporation efficiently depletes strongly FUV-irradiated accretion discs, but it is not sufficient to reliably confirm this conclusion.\\ 
  \textit{Conclusions.} The tools we provided for accurately computing FUV fluxes at the Orion stellar population will be essential for targeting sources in future observations aimed at assessing the role of external photoevaporation on protoplanetary disc.   
  Our study highlights the need for additional measurements of stellar and disc properties across the Orion population, covering the FUV flux range $1-10^{5} \ \mathrm{G}_{0}$.    
  }


   \titlerunning{}
   \authorrunning{Anania, R. et al.}
   \maketitle
   \nolinenumbers

\section{Introduction}
Typical star-forming environments in the Galaxy are illuminated by ultraviolet (UV) radiation emitted by massive stars (mainly of spectral type O and B), that significantly contribute in shaping the physical and chemical properties of circumstellar discs and their planetary systems. Indeed, both numerical models and observations show that protoplanetary discs that are externally irradiated by extreme-ultraviolet (EUV, $\lambda\,<\,912 \ \mathring{\mathrm{A}}$) and far-ultraviolet (FUV, $912 \ \mathring{\mathrm{A}}\,<\,\lambda\,< 2400 \ \mathring{\mathrm{A}}$) photons undergo external photoevaporation of their outermost layers, with the consequent decrease in their mass, size, and lifetime (see \citealt{winter_photoev, Ext_photoevap_review_2025} for two reviews).
In the solar neighbourhood ($\lesssim\,$2 kpc from the Sun), stars are typically subject to FUV fluxes in the range $[10^{2} - 10^{4}] \ \mathrm{G}_{0}$\footnote{$\mathrm{G}_{0}$ is the Habing unit \citep{Habing}, which defines the FUV flux weighted by the average FUV flux in the solar neighbourhood (i.e., $1.6 \times 10^{-3} \ \mathrm{erg \ s^{-1} \ cm^{-2}}$, within 2 kpc to the Sun).} \citep{Fatuzzo_Adams_2008}. 
In this FUV flux range, direct detection of photoevaporative disc winds is challenging as the ionization front is typically excited at $\gtrsim 3\times10^{3} \ \mathrm{G}_{0}$ \citep{Kim_NGC1977} and the peculiar proplyd-like morphology \citep[e.g.,][]{ODell_1993, Ricci_2008} is less apparent. However, observational evidences such as the dependence of the mm-continuum flux emission on the FUV flux \citep{Ansdell_sigmaori, Mauco_sigmaOri, Ansdell_lambda_ori, Soda}, and the gas disc sizes in Upper Scorpius which are explained by the inclusion of external photoevaporation in a viscous model \citep{Anania_AGE_PRO}, 
suggest that external photoevaporation can significantly influence disc properties even under moderate UV radiation fields. These observations are supported by theoretical models showing that dust dynamics and disc chemical properties are affected across a wide range of FUV fluxes \citep[e.g.,][]{Walsh_2013, Sellek_fried, Paine_dust, Keyte_2025}.
However, since 
the star-forming regions located within $\lesssim\,$200 pc from the Sun are weakly FUV-irradiated ($<\,$100$ \ \mathrm{G}_{0}$), 
pursuing the goal of investigating typically FUV-irradiated discs requires observing more distant regions, which introduces complications due to sensitivity and resolution limits, and greater uncertainty in astrometric parameters. In this work, we focus on Orion, which offers a perfect laboratory for studying the relation between young planet birthplaces and the environmental UV radiation, as this is the closest young stellar association ($\sim$360-460 pc, \citeauthor{Orion_dist}\citeyear{Orion_dist}, \citealt{Wright_2020_OB}) hosting late O-type and early B-type stars and active star formation. 
Orion has previously been divided into several macro-areas based on position on the sky, stellar proper motion, and average age \citep[e.g.,]{Kounkel_orion_division_apogee}: Orion A, B, C, and D. The Orion A molecular cloud includes the Orion Nebula Cluster (ONC) and extends to the south-west part of the region. Orion B occupies the north-west part, Orion C corresponds to the central Orion belt, while Orion D spans the north-east region.

Accurately estimating the FUV radiation field is essential to select a representative stellar sample to address the role of the environment on disc properties. The FUV flux at the position of stars might be challenging to evaluate as it strongly depends on the three-dimensional (3D) geometry of a stellar cluster, which is highly influenced by parallax measurements. \citet{Anania_25} proposed using the 2D geometry of a stellar cluster and the argument of isotropy to derive the best estimate 3D separation from OBA-type stars, and then compute the best estimate FUV flux with associated uncertainty. This procedure allowed for accuracy in the FUV flux calculation.
We used the OPTICS clustering algorithm to divide a region of $6^\circ$ radius around the ONC core in sub-clusters, and we applied the method of \citet{Anania_25} to compute the FUV flux at the position of the stars members of the sub-clusters, with the aim of probing the range of FUV flux covered by Orion. We provide a publicly available table of FUV fluxes for a vast stellar population in Orion that will be fundamental to identify the most suitable targets for future works investigating disc evolution in connection with the environment.

We provide insight into the relation between stellar accretion and environmental FUV radiation field in Orion. This investigation is motivated by the fact that it is still poorly understood how the properties and evolution of the innermost disc regions ($\lesssim 10$ au) are linked to the outer disc, which is affected by external photoevaporation and interactions with the environment. Observations show contradictory raesults on the influence of external UV radiation on the chemistry of the inner and outer disc regions \citep[e.g.,][]{Bernè_2024, Frediani_2025}. Theoretical models show that viscous externally UV-irradiated protoplanetary discs tend to evolve toward a constant ratio between the rate of mass lost in photoevaporative winds and that accreting onto the central protostar \citep{Clarke_2007, Winter_sigmaOri, Sellek_fried}, and that the abundance of chemical species in the inner disc region can be significantly affected by external radiation fields \citep[e.g.,][]{Walsh_2013}. 
In this work, we study the stellar accretion luminosities determined from H$_{\alpha}$ emission in \citet{Lavinia_Macc} as a function of the FUV flux and age. We complement this analysis with a population synthesis model of viscous externally irradiated discs to understand whether the stellar accretion luminosity can provide an indication of the effect of external photoevaporation. \\

The work is structured as follows. 
In Sec. \ref{sec:fuv_claculation}, we describe the approach to evaluate the FUV flux at the position of candidate members of Orion. In Sec. \ref{sec:method_models}, we present the sample of stellar accretion luminosities and we describe the  parameters of the disc evolution model. The resulting FUV flux map of Orion is presented in Sec. \ref{sec:results_orion_map}, while the results of the comparison with the distribution of the accretion luminosities are illustrated in Sec. \ref{results:stellar_accretion_vs_fuv}. We discuss the results obtained and the limitations of our method in Sec. \ref{sec:discussion}. Finally, we give our conclusions in Sec. \ref{conclusions}.
\section{FUV flux calculation}\label{sec:fuv_claculation}
We adopted the method detailed in \citet{Anania_25} to infer the FUV flux at the position of a large sample of stars in the Orion region. This approach accounts for parallax uncertainties in the FUV flux calculation by using the 2D geometry of a stellar cluster to make arguments about the 3D separation between target stars and massive stars. We divided the Orion region into sub-clusters, as described in Sec. \ref{sec:sub_clustering}. Subsequently, we evaluated the FUV flux as presented in \ref{subsec:FUV_flux_comp_density}. 
    \subsection{Sub-clustering of Orion}\label{sec:sub_clustering}
    We selected from the Gaia DR3 catalogue all the stars contained in a 6D box defined on the basis of limits on right ascension (RA), declination (Dec), proper motion ($\mu_{\mathrm{RA}}$, $\mu_{\mathrm{Dec}}$), parallax ($\bar{\omega}$), and parallax-over-error ($\bar{\omega}/\sigma_{\bar{\omega}}$). Specifically, we selected stars in a circular region of radius six degrees centred at the $\theta^{1}$C star, which is located in the core of the ONC, the region presenting the highest stellar density in Orion. Then, we limited the proper motion, parallax, and parallax-over-error, considering a large region of the parameter space around the typical ONC values \citep[e.g.,][]{Da_Rio_2012, Kounkel_orion_division_apogee}, which is extended to avoid missing potential stellar members. Specifically, we used $-4 < \mu_{\mathrm{RA}}/\mathrm{(mas \ s^{-1})} < 4$, $-4 < \mu_{\mathrm{Dec}}/\mathrm{(mas \ s^{-1})} < 4$, $2 < \bar{\omega}/\mathrm{mas} < 3.5$, and $\bar{\omega}/\sigma_{\bar{\omega}} > 5$, to constrain the selection of the Orion members.
    Subsequently, we used the OPTICS (Ordering Points To Identify the Clustering Structure, \citeauthor{OPTICS} \citeyear{OPTICS}) clustering algorithm 
    to identify sub-groups of stars, using as clustering criteria the position on the sky (RA, Dec), the proper motion, the parallax for each star, and a minimum number of stars per sub-cluster of 30 \citep{Vioque_2023}. In Sec. \ref{sec:discussion}, we show that requiring 30-40 stars per cluster is sufficient to reliably define the profile of the local density function (i.e., the distribution of pairwise separations or, in case of centrally concentric clusters, the separation of the stars from the centre of the cluster, see Sec. \ref{subsec:FUV_flux_comp_density}).   
    For each star $i$, the algorithm computes a \textit{core distance}, which is the distance to its nearest neighbourhood, and a \textit{reachability distance} from every other star $j$, which is defined as
    \begin{equation}
        \mathrm{reachability}(i,j) = \max( \  core\_distance(i), \ distance(i,j) \ ),
    \end{equation}
    where $core\_distance(i)$ and $distance(i,j)$ must be intended in 6 dimensions because they account for the clustering criteria we imposed (i.e., RA, Dec, $\mu_{\mathrm{RA}}$, $\mu_{\mathrm{Dec}}$, $\bar{\omega}$, $\sigma_{\bar{\omega}}$). The reachability reflects the minimum density level at which $i$ and $j$ belong to the same cluster. By ordering the stars according to their reachability, the algorithm produces a hierarchic plot in which the transitions between clusters that contain at least the minimum number of stars that was initially set (i.e., 30 in our case) are indicated by peaks in the reachability. In total, we isolated 22 main sub-clusters in Orion, as 
    discussed in Sec. \ref{sec:results_orion_map}.

    The work by \citet{Kounkel_orion_division_apogee} has previously investigated the sub-division of Orion into clusters using APOGEE-2 and Gaia DR2, assigning sub-cluster memberships to $\sim5200$ stars. In contrast, our current sample includes $\sim 8600$ stars across 22 sub-clusters. 
    Of the sub-cluster members identified by \citet{Kounkel_orion_division_apogee}, $\sim70\%$ overlap with our sample, and of those, $\sim 96\%$ have sub-cluster memberships consistent with our assignments. Since most of the sub-clusters we found are consistent with those identified by \citet{Kounkel_orion_division_apogee}, in this work we adopt a similar nomenclature for consistency. An important difference between our work and  \citet{Kounkel_orion_division_apogee} is that they divided the main Orion sub-clusters into even smaller sub-groups containing a number of members ranging from 10 to $\sim300$, leading to a total of 190 sub-groups. Instead, we required a minimum of 30 stars per cluster in order to assign reliable density distribution functions and compute FUV fluxes (see Sec. \ref{subsec:FUV_flux_comp_density} and \ref{subsec:discussion_limitations}), which resulted in 22 main sub-clusters spanning the Orion area. 
    In general, the sub-clusters identified in this work contain a number of members between $\sim$60 and $\sim$3000 and have a diameter that spans a few degrees in the sky. \\

    An important consideration on the division in sub-clusters presented in this work is that the stellar membership of the low-mass stars (i.e., types later than A) is not complete, as Gaia DR3 may miss some of the faintest stars (e.g., those embedded in dense cloud cores such as NGC 2024), and the low-mass stars (i.e., G $\gtrsim21$ mag, which is the limit magnitude of Gaia and corresponds to $M_{\star} \lesssim 0.1$ at 400 pc). However, cross-matching our final sample with IR-based catalogues of YSOs \citep{Megeath_Spitzer_orion, vision_vista_orion, NEMESIS_orion}, we verified that $\sim72\%$ of the sample in the IR-based catalogues associated with the Orion region has Gaia DR3 counterpart, and therefore we can conclude that Gaia DR3 has good level of completeness in Orion. 
    Moreover, we highlight that the main goal of this work is not to provide complete and definitive memberships to the entire Orion population. Instead, the level of completeness that we need for our calculation is related to the ability to define a reliable shape of the 2D density distribution function of the region, to compute the FUV flux. As will be shown in Sec. \ref{subsec:discussion_limitations}, 
    setting the minimum number of stars per cluster to 30-40 allows us to recover the density distribution function profile, and thus evaluate the FUV flux. A future addition of stars to the sub-clusters will not significantly influence the shape of the density function, and therefore will negligibly impact our FUV flux calculation (see also Appendix \ref{appendix:incomplete_membership}).    
      
    Our aim is computing the FUV flux of a large population of Orion members, and therefore it is essential to locate the OBA-type stars responsible for the UV radiation and define which of them are members of the identified Orion sub-clusters or field stars. We gained information on the massive stars in the region by combining multiple stellar catalogues in order to compensate for the limitations of each catalogue. In particular, following the approach in \citet{Anania_25}, we combined the ALSIII catalogue of massive stars \citep{ALSIII}, from which we selected all the most massive stars in 2 kpc from the Sun, the Gaia DR3 Extended Stellar Parametrizer for Hot Stars (ESP-HS, \citealt{GaiaDR3_ESPHS, GaiaDR3_ESPHS_2}) catalogue, and \citet{Zari_2021} catalogue of hot stars where we selected only those with effective temperature $>9000$ K (i.e., from $\sim$A0 to earlier spectral types). Moreover, we included the OB stars that are not present in Gaia, because of their high luminosity, but are instead listed in Hipparcos \citet{Hipparcos_catalogue}. Finally, we cross-matched the selected sample of massive stars with the OB stars in Orion listed in \cite{Quintana_Wright_OB} to add the stars that may be missing from our previous selection. 
    The final list of massive stars includes known multiple systems and the FUV flux calculation accounts for their luminosities. Since the population of the most massive OB stars in Orion has been studied extensively, the most massive members are well identified, as well as their multiple companions. Therefore, any correction to the total FUV flux from missing OB companions is unlikely to exceed a factor of $\sim2$.
    In order to derive the FUV luminosity of the OBA stars, we first derived their effective temperature from either spectral type classification \citet{GOSC_catalogue, Gray_Corbally_2009} or the values of effective temperature contained in the ESP-HS Gaia DR3 pipeline. We used these effective temperatures, and the stellar parameters evaluated from MIST isochrones \citep{Mist_1, Mist_2}, in the stellar atmosphere models ATLAS9 \citep{Castelli_Kurucz_2004} to compute the stellar luminosities and integrate them in the FUV range of wavelengths. \\

    In conclusion, we selected 22 main sub-clusters located in correspondence to the Orion A and B molecular clouds and spanning Orion C (central belt), Orion D (north-east), and clusters in the extended population of Orion (see Sec. \ref{sec:results_orion_map}, Fig. \ref{fig:Orion_map_sub_G0}).
    We found $\sim 115$ OBA stars classified as members of the sub-clusters. 
    In our calculation, we also included the contribution of the field OBA-type stars to the total FUV flux, following the approach detailed in the next Section.   

    \subsection{Computing the FUV flux using the local density function}\label{subsec:FUV_flux_comp_density}
    As described in \citet{Anania_25}, the FUV flux in a given location $p$ is computed by summing up the individual FUV fluxes induced by the surrounding OBA stars: $
    F_{\mathrm{FUV, p}} = \sum_{m} (L_{\mathrm{FUV,m}}/{4 \pi |x_{\mathrm{p}} - x_{\mathrm{m}}|^{2}})$, where the separation from massive stars in 3D space, $|x_{\mathrm{p}} - x_{\mathrm{m}}|$, is the main source of uncertainty due to parallax measurements. Despite the fact that Orion is a relatively close star-forming association ($\sim$400 pc from the Sun), astrometric measurements of low-mass stars, which may appear faint, or very bright stars, which may saturate, might be challenging, resulting in a very uncertain (or not defined) parallax. Moreover, the FUV flux tends to be underestimated if the uncertainty in the radial distance of a certain star is larger than the relative separation between that star and the nearest OBA stars.  
    We addressed this problem using the information on the 2D geometry of the stellar clusters in Orion to infer, for each star, the 3D separation from the OBA stars members of the clusters. 
    Specifically, for the stellar clusters  presenting a substructured 2D geometry, we define the probability of finding a pair of stars $ij$ separated by a 3D distance $r_{ij}$, given their 2D projected separation in the sky plane $R_{ij}$:
    \begin{equation}
        \mathrm{d} P (r_{ij} | R_{ij}) = \mathrm{d} P (R_{ij} | r_{ij}) \ \mathrm{d} P(r_{ij}) = \begin{cases} \frac{ 4 \pi R_{ij} r_{ij}}{\sqrt{r_{ij}^2 - R_{ij}^2}}\hat{\rho}_\mathrm{pairs} ( r_{ij})  & R_{ij} < r_{ij} \\ 0 & \mathrm{otherwise}
    \end{cases},
    \label{eq:prob_rR_final}
    \end{equation}
    where $\hat{\rho}_{\mathrm{pairs}} (r)$ is the normalized 3D local density function of the cluster (i.e., the normalised 3D pairwise distribution). Assuming isotropy in the neighbourhoods of each star, $\hat{\rho}_{\mathrm{pairs}} (r)$ is derived from the normalized 2D local density function $\hat{\Sigma}_{\mathrm{pairs}} (R)$ using the Abel's inversion \citep{Abel_1826}:
    \begin{equation}
        \hat{\rho}_{\mathrm{pairs}} (r_{}) = - \frac{1}{\pi} \int_{r}^{\infty} \frac{ \mathrm{d} \hat{\Sigma}_{\mathrm{pairs}} (R) }{ \mathrm{d} R} \frac{1}{\sqrt{ r^{2} - R^{2}}} \mathrm{d} R ,
    \label{eq:Abel_rho}
    \end{equation}
    which is valid if $\hat{\rho}_{\mathrm{pairs}} (r) \rightarrow 0$ as $r \rightarrow \infty$, and where $\hat{\Sigma}_{\mathrm{pairs}} (R)$.
    For the Orion sub-clusters showing a centrally concentric 2D geometry, we employed the assumption of spherical symmetry and the same expressions in Eq. \eqref{eq:prob_rR_final} and Eq. \eqref{eq:Abel_rho} where $R$ and $r$ are the 2D and 3D distances of the stars from the centre of the cluster. Finally, we used the best estimate 3D separation to compute the FUV flux at the position of each star and its associated uncertainty. 
    We added to the FUV flux at each position the contribution from field OBA stars that are not cluster members but are located within the region of interest. Since the local density function method cannot be applied to non-members, we estimated their 3D separations by sampling from the normal posterior distribution of each stellar parallax, adopting a width defined by the parallax uncertainty reported in Gaia DR3 (or Hipparcos, in cases where a Gaia parallax is unavailable).
    Using this approach, we build an FUV flux map of the Orion region (see Sec. \ref{sec:results_orion_map} and Fig. \ref{fig:Orion_map_sub_G0}). In Appendix \ref{appendix:table_fuv}, we include part of the list of FUV fluxes that is available at the CDS, while in Appendix \ref{appendix:sut_comparison_maps} we compare our approach for estimating the FUV flux with a previous method based on dust emission maps, and we discuss the advantages of our approach.

\section{Stellar accretion dataset and model}\label{sec:method_models}
    We aim to explore the relation between stellar accretion and external FUV radiation field in Orion and investigate whether the accretion luminosity could be a useful parameter to test the impact of external photoevaporation on protoplanetary discs. In this Section we present the data sample and the disc evolution model used. 
    \subsection{Accretion luminosity measurements}
    In this work, we used the stellar accretion luminosities from \citet{Lavinia_Macc}, which were derived from Gaia DR3 XP spectra \citep{De_angeli_gaiaXP_2023} of young stellar objects (YSOs) located within 500 pc of the Sun that show H$_{\alpha}$ line emission.
    Although stellar accretion properties derived from H$_{\alpha}$ emission in Gaia-based observations are affected by larger uncertainties compared to spectroscopic measurements, the sample of \citet{Lavinia_Macc} spans the entire Orion region and allows us to explore a wide range of FUV fluxes.
    Stellar accretion luminosities and mass accretion rates in Orion have previously been derived spectroscopically for 50 stars in the $\sigma$ Ori region \citep{Mauco_sigmaOri} and 37 stars in the Orion Nebula \citep{Piscarreta_2025} using the ESO VLT/X-Shooter spectrograph \citep{XShooter}. This sample of spectroscopic measurements includes a small number of objects that span only two of the 22 Orion sub-clusters identified in Sec. \ref{sec:results_orion_map}. In Appendix \ref{appendix:compare_gaia_xshooter_lacc}, we show and discuss the comparison between the accretion luminosities of the sample of objects that have both X-Shooter spectra and H$_{\alpha}$ emission in Gaia XP spectra.

    We found that 1200 candidate members of Orion for which we computed the FUV flux have $H_{\alpha}$ line emission in Gaia DR3 XP spectra and a defined accretion luminosity in \citet{Lavinia_Macc}. An additional sample of 75 objects presents H$_{\alpha}$ emission compatible with chromospheric activity, so we considered their accretion luminosities as upper limits.
    To properly investigate the relation between stellar accretion and FUV radiation field at the population level, it is necessary to account for observational incompleteness by including the detection fraction in the calculation. Indeed, neglecting non-detections would bias the inferred trends and lead to an incomplete characterization of the underlying relation between stellar accretion and external radiation. 
    However, non-detections in Gaia DR3 XP spectra can result from multiple physical and observational effects, which vary between sources, making it difficult to distinguish genuinely non-accreting sources from accreting but undetected objects (see \citealt{Lavinia_Macc} and Sec. \ref{subsec:discussion_accretion_rate} for a detailed discussion). Therefore, determining a representative selection function for the entire investigated sample is challenging.
    We addressed this problem by considering the population of 5073 objects in Orion that have Gaia DR3 XP spectra but no H$_{\alpha}$ emission is reported in \citet{Lavinia_Macc}. For this sample of objects, we defined representative accretion luminosities by modelling the intrinsic relation between accretion luminosities and stellar luminosities, and by making well-founded assumptions on the detection probability. Specifically, we used the measured stellar luminosities and their associated observational uncertainties, and adopted a linear model with intrinsic scatter for the true (unknown) accretion luminosities.
    We then constructed a parametric detection curve in accretion luminosity, which is a function of stellar luminosity and allows us to define limit accretion luminosities, of the kind:
    \begin{equation}
        L_{\mathrm{acc,lim}}(L_\star) = L_{\mathrm{acc,0}} + f\,L_\star\left[1+\left(\frac{L_\star}{L_{\rm thr}}\right)^2\right],
    \end{equation}
    where $L_{\mathrm{acc,0}}$ and $L_{\rm thr}$, in units of $L_\odot$, and the dimensionless $f$ are treated as free parameters with associated prior distributions. 
    The complete procedure is detailed in Appendix \ref{appendix:lacc_upp}.
    For consistency, we applied the same modelled detection curve to the population synthesis model presented in the following Section.
     
    This list of objects provides us with a homogeneous sample of YSOs to study the relation between stellar accretion and external FUV radiation field.
    Although \citet{Lavinia_Macc} reported both accretion luminosities and mass accretion rates of the stellar sample, we decided to compare FUV fluxes with accretion luminosities. This choice is motivated by the fact that accretion rate, $\dot{M}_{\mathrm{acc}}$, is derived from accretion luminosity, $L_{\mathrm{acc}}$, and stellar parameters via the standard relation
    \begin{equation}
        \dot{M}_{\mathrm{acc}} =\frac{L_{\mathrm{acc}} R_{\star}}{G M_{\star}}\left( 1 - \frac{R_{\star}}{R_{\mathrm{in}}} \right)^{-1},
        \label{eq:accretion_rate_lum}
    \end{equation}
    where $M_{\star}$ and $R_{\star}$ are the stellar mass and radius, respectively, and $R_{\mathrm{in}}$ is the inner disc radius. Since these stellar parameters are typically inferred from stellar evolutionary isochrones and are strongly affected by extinction, which is uncertain in Orion, we opted to base our analysis on $L_{\mathrm{acc}}$, which is less affected by extinction uncertainties.

    \subsection{Disc evolution model}\label{subsec:disc_models_parameter_space}
    We complemented our analysis with a population synthesis model of protoplanetary discs subject to viscous evolution and external photoevaporation.
    We updated the population synthesis code Diskpop \citep{Somigliana_diskpop} to include the process of external photoevaporation. Specifically, we assumed that each disc of the population is subject to a constant FUV flux, and we define the mass loss rate due to external photoevaporation by interpolating the FRIEDv2 grid of mass loss rates \citep{FRIEDv2} based on the stellar and disc parameters. Disc material is removed outside-in following the numerical approach of \citet{Sellek_fried}. 
    We modelled a population of geometrically thin and axisymmetric discs undergoing viscous evolution \citep{LeP_solution} and external photoevaporation, where the time evolution of the disc surface density profile is described by
    \begin{equation}
        \frac{\partial}{\partial t} \Sigma_{\mathrm{g}} = \frac{3}{R} \frac{\partial}{\partial R} \left( R^{1/2} \frac{\partial}{\partial R} (\nu \Sigma_{\mathrm{g}} R^{1/2})  \right) - \dot{\Sigma}_{\mathrm{ext}},
    \label{eq:diffusion_eq}
    \end{equation}
    where $\nu = \alpha c_{\rm{s}} H_{\rm{g}}$ is the kinematic viscosity, depending on the gas scale height $H_{\rm{g}}$ and the sound speed $c_{\rm{s}}$, $\alpha$ is the \citet{alpha_prescription} parameter describing the disc viscosity, and $\dot{\Sigma}_{\mathrm{ext}}$ is the mass loss rate due to external photoevaporation.  
    Assuming that the various sub-clusters in Orion have a similar initial mass function (IMF), we modelled a population of 7500 objects with stellar masses randomly selected from a Kroupa IMF \citep{Kroupa_2001} in the range $[0.1 - 3] \ \mathrm{M}_{\odot}$, which is the range of stellar masses supported by the FRIEDv2 grid. We applied a cut to the sample of stellar masses considering an average extinction based on that observed in Orion \citep{DustExtinction_map}, and a limit in G magnitude based on Gaia XP spectra, i.e. we excluded objects with G magnitude higher than 17.65, which corresponds to stellar masses smaller than $\sim 0.2-0.3 \ \mathrm{M}_{\odot}$ at the distance of Orion. The final sample consists of $\sim 7400$ stars with disc that we modelled using the parameter space of the initial conditions presented in Table \ref{table:models_population}.
    Specifically, the FUV fluxes are randomly selected from the distribution of FUV fluxes of Orion derived in this work. The initial disc mass and the initial disc extent are set to representative values of 5$\%$ of the stellar mass, and 50 au, respectively. We show in Appendix \ref{appendix:lacc_fuv_mr} that varying the initial disc mass and extent only leads to a shift in the accretion luminosity, without significant consequences on the relation with the FUV flux, which is the relation we are focusing on. Moreover, we applied to the accretion luminosities retrieved from our models a spread consistent with the observational uncertainty on Gaia-based $L_{\mathrm{acc}}$, which is larger than the scatter caused by varying the initial disc mass and the characteristic radius in our model.   
    The condition imposed on the $\alpha$ viscosity parameter ensures to achieve the commonly observed relation between accretion and stellar mass $\dot{M}_{\mathrm{acc}} \propto M_{\star}^{2}$ \citep{Alexander_Armitage}.
    We evolved the disc population up to 10 Myr, which is the average age of the old Orion stellar population \citep[e.g.,][]{Kounkel_orion_division_apogee}. The comparison between observations and the population synthesis model is presented in Sec. \ref{results:stellar_accretion_vs_fuv}.
    \begin{table}[h!]
    \caption{Parameter space used for population synthesis}
    \centering
    \begin{tabular}{c|c}
    \hline\hline
    Parameter & Values \\
    \hline
    $M_{\star} \ (\mathrm{M}_{\odot})$ & Kroupa IMF [0.1 - 3]\\
    $M_{\mathrm{d,0}} \ (\mathrm{M}_{\odot})$ & 0.05 $M_{\star}$ \\
    $R_{\mathrm{c,0}} \ (\mathrm{au})$ & 50\\
    $\alpha$ & $10^{-3}(M_{\star}/\mathrm{M}_{\odot}$)\\
    FUV flux (G$_{0}$) & sampling [1-$10^{7}$] \\
    \hline
    \end{tabular}
    \tablefoot{We modelled a population of $\sim 7500$ stars with disc, where the stellar masses are drawn from a Kroupa IMF in the range [0.1 - 3] $\mathrm{M}_{\odot}$, and the FUV fluxes are randomly selected from the distribution of Orion FUV fluxes computed in this work. The initial disc mass ($M_{\mathrm{d,0}}$) is set to 5\% of the stellar mass, the initial disc characteristic radius ($R_{\mathrm{c,0}}$) is fixed to 50 au, and the $\alpha$ viscosity parameter increases linearly with stellar mass.}
    \label{table:models_population}
    \end{table}

\section{Results}
\subsection{Orion FUV flux map}\label{sec:results_orion_map}
\begin{figure*}[h!]
    \centering
    \includegraphics[width=0.43\textwidth]{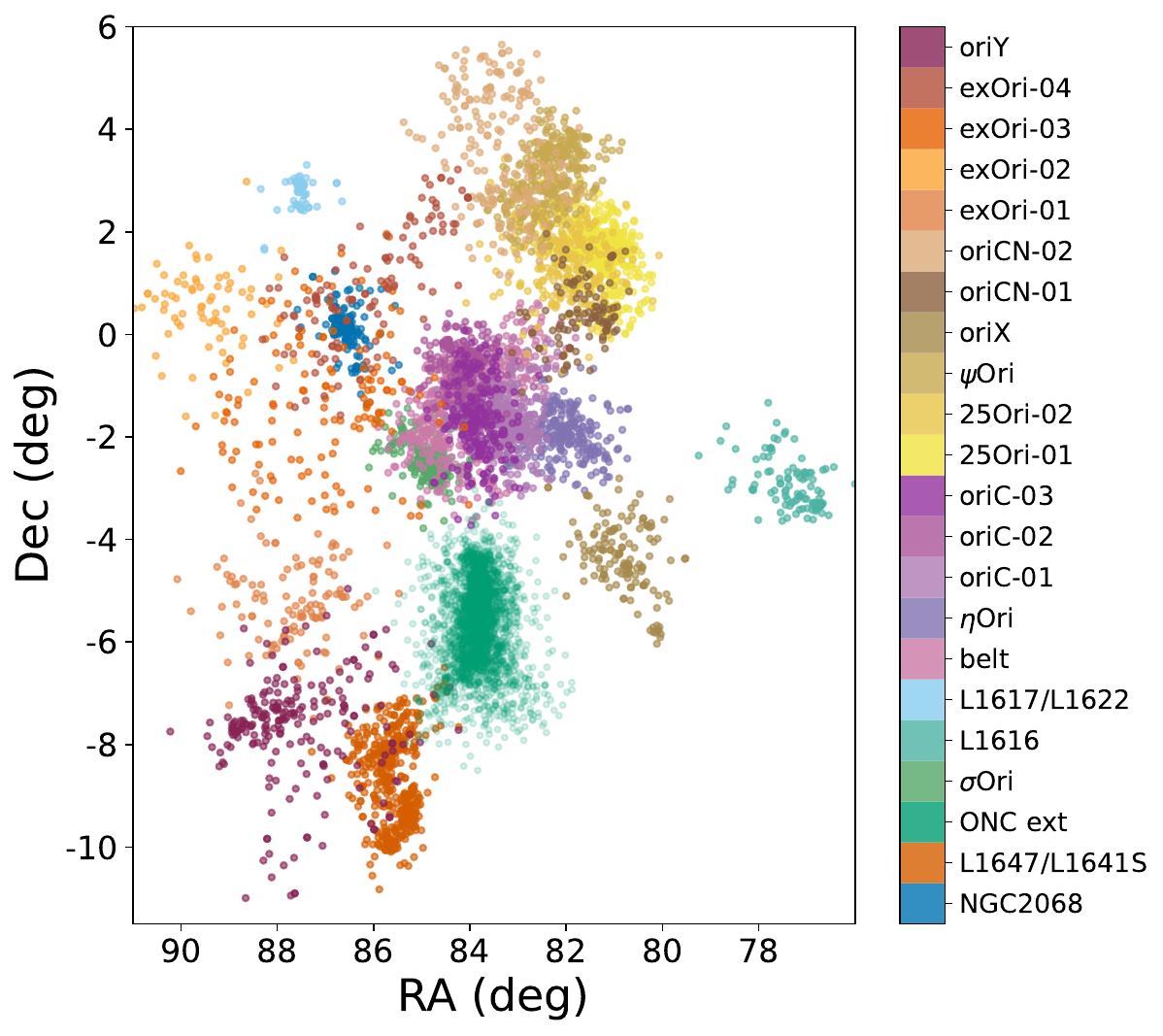}
    \hspace{1.5cm}
    \includegraphics[width=0.394\textwidth]{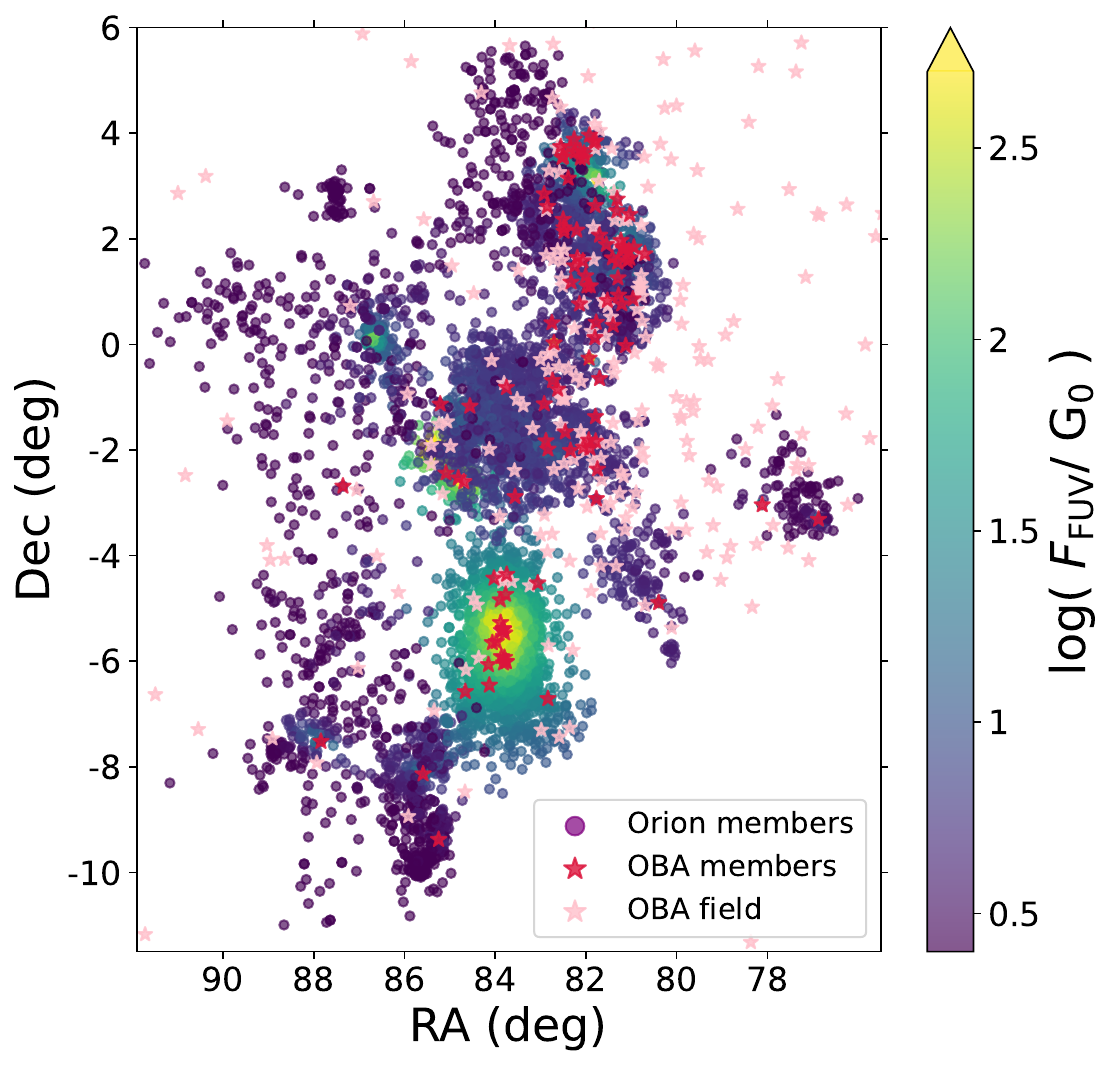}
    \caption{\textit{Left panel:} The main 22 Orion sub-clusters obtained using the OPTICS clustering algorithm (Sec. \ref{sec:sub_clustering}). The sub-clusters are ordered in the colorbar from bottom to top by increasing average age, from $<1$ Myr to $>6$ Myr.  \textit{Right panel:} Median FUV flux at the position of the Orion members identified in this work. The full table of FUV flux values, including uncertainties on the FUV flux, is available at the CDS (see Appendix \ref{appendix:table_fuv}, Table \ref{appendix:table_table_fuv}). Red and pink points mark the OBA members and field stars, respectively.}
    \label{fig:Orion_map_sub_G0}
\end{figure*}
\begin{figure}
    \centering
    \includegraphics[width=0.47\textwidth]{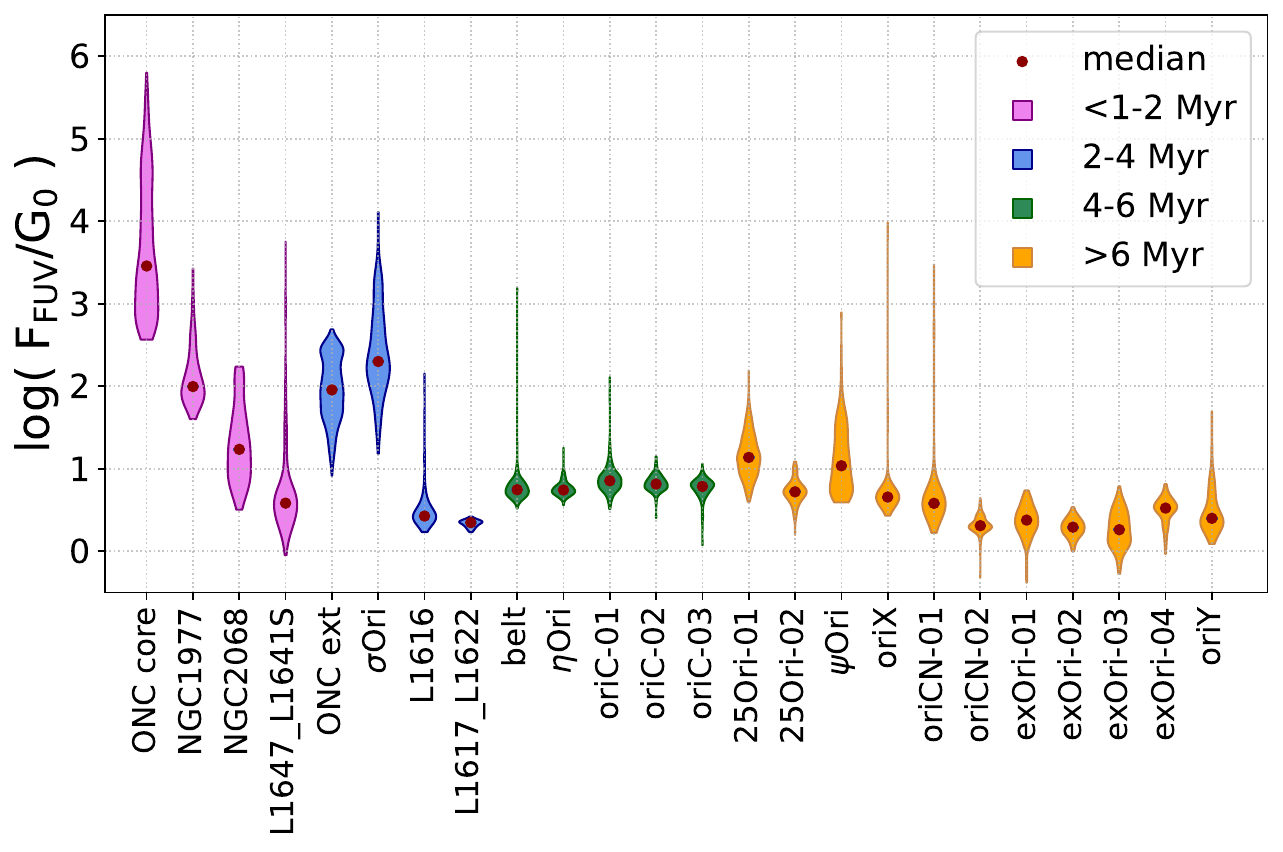}
    \caption{Distribution of the median FUV flux for the stellar sample identified in this work in the Orion sub-clusters. x-axis: Commonly known name of the sub-clusters identified in this work, ordered by age from left to right. Notes: We isolated here the ONC core ($\sim$3 pc radius from $\theta^{1}$C) and we included the FUV fluxes for the NGC1977 sources investigated in \citet{Anania_25}.}
    \label{fig:violin}
    \end{figure}
In this Section, we show the FUV flux values resulting from applying the density distribution function method described in Sec. \ref{subsec:FUV_flux_comp_density} at the position of the Orion members in each sub-cluster.
The complete table of FUV fluxes (and their uncertainties) is publicly available at the CDS. We also provide a code to compute the FUV flux at the position of additional stars in the Orion sub-clusters (see details below), which is available on Github (see Data availability Section).

Using the OPTICS algorithm, we identified 22 sub-clusters in Orion, as shown in the left panel of Fig. \ref{fig:Orion_map_sub_G0}. The geometry of the clusters defines the shape of the normalised density distribution function that we used to ultimately compute the FUV flux.
The right panel in Fig. \ref{fig:Orion_map_sub_G0} shows the median FUV flux at each stellar object, resulting from applying the method described in Sec. \ref{subsec:FUV_flux_comp_density}. The locations of the OBA stars, members of the clusters and field stars, are marked in red and pink, respectively.
The distribution of the median FUV fluxes for each sub-cluster is presented in Fig. \ref{fig:violin}, where we isolated the ONC inner core (i.e., the Trapezium, which hosts the most massive stars in the region) and NGC 1977 from the extended ONC region. This is done in order to extract these two small clusters from the full ONC identified by OPTICS (see below).
We highlight here the main properties and the age range \citep[for the age estimate see e.g.,][]{Warren_Hesser_1978, Briceno_2005_ob_age, Kounkel_orion_division_apogee} of the sub-clusters:
\begin{itemize}
    \item The ONC region includes its innermost core (the Trapezium), snd also NGC 1977.
    Despite of the astrometric accuracy of the Gaia DR3 data, the uncertainties in parallaxes and proper motions for stars in the densest stellar region like the ONC are still too high to allow a solid separation of the ONC in kinematically distinct sub-populations.
    Therefore, we treated the ONC as a cluster containing multiple substructures and defined the corresponding local density function that describes its geometry (see \citealt{Anania_25} for the detailed numerical approach related to this geometry). We only isolated the ONC core ($\sim 0.3$ deg from $\theta^{1}$C) and NGC 1977, which are younger ($<2$ Myr, e.g., \citealt{Beccari_2017}) than the extended ONC stellar population, and therefore are treated with their own normalised density function. In Fig. \ref{fig:violin}, the FUV fluxes of the sources located in the innermost ONC (\citealt{Ricci_2008}, \citealt{Eisner_2018}, ONC core in Fig. \ref{fig:violin}), and in NGC 1977 (Kim et al., in prep.) are taken from \citet{Anania_25}, where a more detailed study on these regions was performed. 

    \item The regions presenting young ages of $\lesssim1-2$ Myr are: the ONC core and L1647/L1641-South, which are located in the Orion A cloud, NGC 2068 and NGC 2024, which are part of the Orion B cloud.

    \item The outer part of the ONC is known to be older than its core, with an age range $\sim 2 - 3$ Myr. A similar age is associated with L1616, which is located in the eastern part of Orion, and the cometary clouds L1617 and L1622.
    While we identified stars located in the vicinity of the clouds, the sources showing infrared excess and associated with L1622 reported by \citet{L1617_L1622} are not included in our sample, as they lack Gaia DR3 identifier or exhibit discrepant astrometric properties, such as higher proper motions and/or smaller Gaia DR3 distance estimates.  

    \item The $\sigma$ Ori cluster is in projection closer to Orion OB1b. However, its age is more similar to that of the outer part of the ONC (i.e., Orion OB1c), which is $\sim 3-4$ Myr \citep{Jeffries_2006_sigmaOri, Hernandez_2014_sigmaOri}. The kinematic association of this region to Orion OB1b is still debated \citep[e.g.,][]{Caballero_solano_sigmaOri, Jeffries_2006_sigmaOri}. In our division of Orion, NGC 2024 and $\sigma$ Ori fall within the same sub-cluster, as the former is still embedded and the stellar astrometry in this region remains highly uncertain, leaving only a few stars that can be reliably assigned as members (i.e., not enough to assign a density function of the region). Nevertheless, we emphasize that NGC 2024 is very young, with an age comparable to NGC 2068 and the ONC core.  

    \item The sub-cluster named after \textit{belt} in Fig. \ref{fig:Orion_map_sub_G0} and Fig. \ref{fig:violin} spans the Orion belt in correspondence of the stars $\zeta$ Ori, $\epsilon$ Ori, and $\delta$ Ori. Together with the clusters named after $\eta$ Ori and Ori C -$[01 - 02- 03]$, this region defines the central part of Orion corresponding to Orion OB1b. The average age of these clusters spans $\sim 4 - 6$ Myr \citep{Briceno_2005_ob_age, Calvet_2005, Bally_2008}.  

    \item The sub-clusters 25 Ori-$[01-02]$, $\psi$ Ori, and Ori CN-$[01-02]$ are located in the north-eastern part of Orion, in the Orion OB1a region, where there is little or no gas left. This region hosts the oldest stellar population with average age of $\sim 7 - 10$ Myr \citep{Briceno_2005_ob_age, Calvet_2005}.

    \item The remaining clusters form the extended population of Orion, which is of similar age to Orion OB1a, $\sim 7-10$ Myr \citep{Briceno_2005_ob_age}.
\end{itemize}

We emphasise that the upper limits of the FUV flux distributions presented in Fig. \ref{fig:violin} may change in the future if following studies will identify candidate members closely located to the most massive stars, which may not be included in our list due to their uncertain astrometry because of the proximity to luminous stars. 
Since we are aware of the fact that future studies and different techniques may lead to improvements of the stellar membership, and that we may need to compute the FUV flux at the position of additional stars, we released a \texttt{Python} script that, given the main astrometric parameter of a star (RA, Dec, $\mu_{\mathrm{RA}}$, $\mu_{\mathrm{Dec}}$, $\bar{\omega}$, $\sigma_{\bar{\omega}})$, provides the probability that the star is associated with any of the 22 identified clusters in Orion. This routine returns the FUV flux and its uncertainty as $16^{\mathrm{th}}$, $50^{\mathrm{th}}$, and $84^{\mathrm{th}}$ percentiles of the posterior distribution. The code is structured to accept as input a single star or multiple stars at the same time.  We emphasise that this FUV flux calculation routine is built on the sub-clustering division of Orion that is made in this work, and derives the 3D separation of target stars from OB stars by making arguments about the sub-cluster geometry. Therefore, it may be subject to future updates following major revisions on the sub-cluster membership estimates.
See Sec. \ref{subsec:discussion_limitations} for a discussion on the cluster memberships and the limitations of the method.\\

The FUV flux in young and more nearby regions (within $\sim200$ pc) such as Lupus and Taurus does not exceed $\sim 10 \ \mathrm{G}_{0}$ \citep{Anania_25}, while Orion is the closest region that also hosts a population of young ($\lesssim6$ Myr), intermediately irradiated stars, with FUV flux in the $\sim 10^{2} - 10^{4} \ \mathrm{G}_{0}$ range (see right panel of Fig. \ref{fig:Orion_map_sub_G0}). This population consists of $\sim35\%$ of the total Orion sample investigated in this work. 
Therefore, although the intermediately irradiated population does not dominate the overall FUV flux distribution in Orion, this region nonetheless hosts a population of key potential candidate targets in this FUV flux range that is absent in more nearby regions.
In contrast, our study demonstrates that Orion is not the ideal region for investigating the effect of extreme UV radiation on a population of protoplanetary discs. Indeed, the sample of stars illuminated by extreme FUV radiation fields (i.e., $\gtrsim 10^{4} \ \mathrm{G}_{0}$), including proplyds, consists of only $\sim5\%$ of the stellar population investigated in this work. To study disc evolution under extreme environmental conditions, more distant regions such as Carina, M16, and M17, that host a population of early O-type stars, are better suited.\\

The majority of the Orion stellar population investigated in this work, $\sim 68\%$, is illuminated by FUV fluxes in the range $1-10^{2} \ \mathrm{G}_{0}$ (see Fig. \ref{fig:violin}). 
Most of these sub-clusters are located in the north-east region of Orion (Orion OB1a). Although this region hosts numerous B-type stars ($\sim$55), it presents a lower stellar density due to its older age, resulting in greater average distances from the most massive stars compared to the ONC. As a result, even though the UV radiation levels are comparable to those in more nearby regions ($<$200 pc), the weakly irradiated areas of Orion are in general older ($>6$ Myr) and located in an OBA-rich environment, where cluster dynamics may have significantly influenced the FUV radiation history and the evolution of protoplanetary discs.
In Orion OB1a, 25 Orionis shows moderate FUV radiation fields of $\sim 10-100 \ \mathrm{G}_{0}$ and evolved YSOs ($\sim 5-10$ Myr), making it comparable to Upper Scorpius. However, the stellar structure of the two regions is different: 25 Orionis is illuminated by a central B1V star (25 Orionis itself) and a sparse population of late-type B stars, whereas Upper Scorpius presents a sub-structured population around many late-type B stars \citep{Anania_AGE_PRO}. A future comparison between these two stellar populations could offer insight into the role of the stellar environment on disc evolution. 

\subsection{Stellar accretion vs external FUV radiation field}\label{results:stellar_accretion_vs_fuv}
In this Section, we analyse the relation between stellar accretion luminosity inferred from H$_{\alpha}$ line emission in Gaia XP spectra and FUV radiation field in Orion, and compare the outcomes with numerical disc evolution models.
    \begin{figure*}
        \centering
        \includegraphics[width=0.98\textwidth]{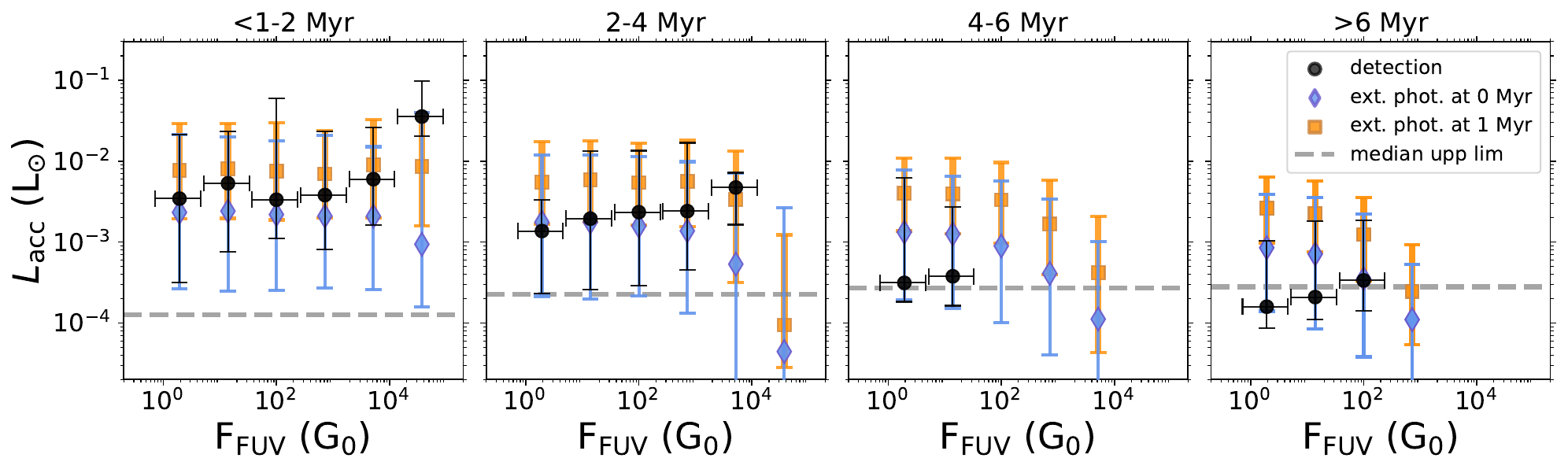}
        \caption{The markers indicate the median accretion luminosity $L_{\mathrm{acc}}$ in each bin of FUV flux as derived from the Orion objects in \citet{Lavinia_Macc} (black dots) and predicted by the disc population synthesis model where external photoevaporation acts from the beginning (blue diamonds) and after 1 Myr to mimic the effect of extinction (orange squares). The vertical error bars extend from the 16$^{\mathrm{th}}$ to the 84$^{\mathrm{th}}$ percentile or the distributions, while the horizontal bars indicate the width of the FUV flux bins. The four panels refer to four increasing age bins from left to right: We divided the observed disc sample by average cluster age, the results of the model are taken at 1 Myr, 3 Myr, 5 Myr, and 10 Myr, respectively. Dashed lines indicate the median upper limit $L_{\mathrm{acc}}$ observed in each age bin.}
    \label{fig:histo_lacc_fuv}
    \end{figure*}
    \begin{figure*}
        \centering
        \includegraphics[width=0.98\textwidth]{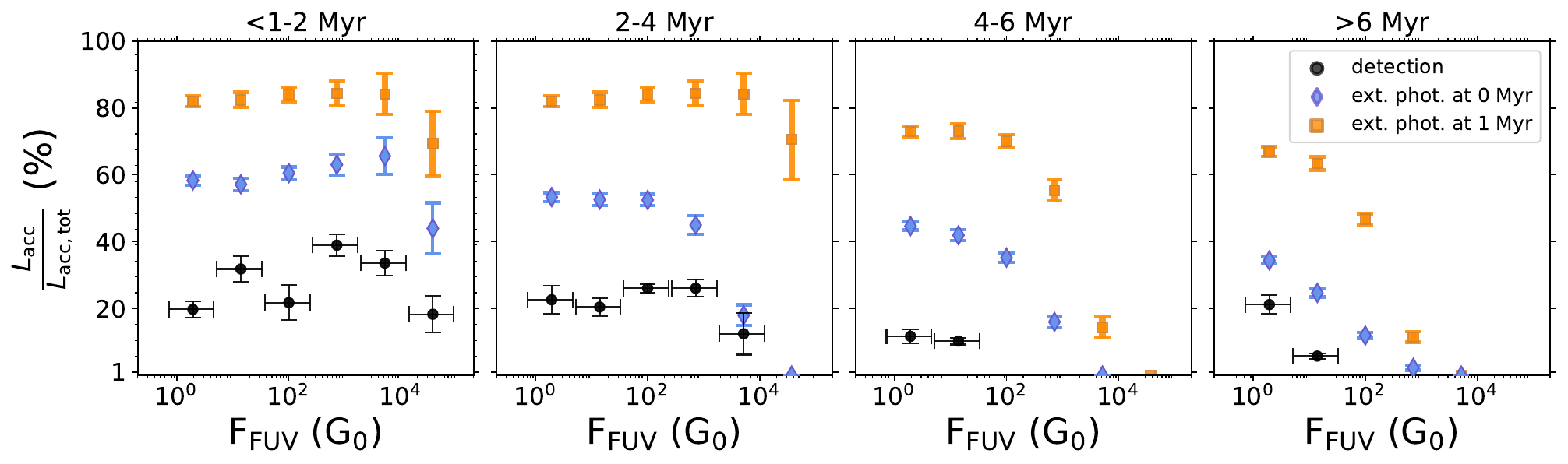}
        \caption{Detection fraction of  $L_{\mathrm{acc}}$ per bin FUV flux and increasing age from the left to the right panel. Black dots refer to the observations, while blue diamonds and orange squares are the results of the population synthesis model where external photoevaporation is included from the beginning of the evolution and after 1 Myr (to mimic the effect of extinction), respectively. The vertical error bars are the uncertainties considering a binomial distribution, while the horizontal bars indicate the width of the FUV flux bins. The age of the observed sample refer to the average age of the sub-clusters, while the results from the model are taken at 1 Myr, 3 Myr, 5 Myr, and 10 Myr, respectively from the left to the right panel.}
        \label{fig:histo_fraction}
    \end{figure*}  
    Figure \ref{fig:histo_lacc_fuv} presents the median $L_{\mathrm{acc}}$ in each FUV flux bin for the Orion sample investigated in this work (black dots) and the results of the population synthesis models (blue diamonds and orange squares). 
    The panels in Fig. \ref{fig:histo_lacc_fuv} illustrate the results in four different age bins, from the youngest (left) to the oldest (right). Specifically, the Orion population is divided into the following age bins (see also Fig. \ref{fig:violin}, and for references on age range see \citealt{Warren_Hesser_1978, Briceno_2005_ob_age, Kounkel_orion_division_apogee}).\vspace{0.1cm}
    
    \noindent $\bullet$ $\lesssim 1-2$ Myr : stars in the ONC core (within 0.3 deg from $\theta^{1}$C), L1647/L1641-S, and NGC 2068;\vspace{0.2cm}
    
    \noindent $\bullet$ $2-4$ Myr : stellar population of $\sigma$ Ori, L1616, and L1617/L1622;\vspace{0.2cm}
        
    \noindent $\bullet$ $4-6$ Myr : stellar sample in belt, $\eta$ Ori, and Ori C-[01, 02, 03];\vspace{0.2cm}
        
    \noindent $\bullet$ $>6$ Myr : the remaining clusters in Orion OB1a and the extended population of Orion.\\
    
    The population synthesis consists of 1D viscous evolution models of gaseous protoplanetary discs where the effect of external photoevaporation is included, as described in Sec. \ref{subsec:disc_models_parameter_space}. The initial conditions of the disc population are presented in Table \ref{table:models_population}.
    We performed two types of population synthesis model. In the first model, external photoevaporation acts from the beginning of disc evolution (blue markers in Fig. \ref{fig:histo_lacc_fuv}), while in the second model we introduce external photoevaporation after 1 Myr of evolution to mimic the effect of dust extinction (orange markers in Fig. \ref{fig:histo_lacc_fuv}). Indeed, dust extinction is capable of shielding protoplanetary discs in the first stages of their lifetime. This is motivated by simulations of molecular clouds, which show that dust shields efficiently protoplanetary discs against external FUV radiation during the first $\sim 0.5-1$ Myr, whereas the effectiveness of this shielding decreases sharply after that time \citep[e.g.,][]{Ali_shield, Qiao_extinction_2022}. 
    In our models, the accretion luminosity is evaluated, for each object at each timestep, using the relation in Eq. \eqref{eq:accretion_rate_lum}, where the mass accretion rate is an output of the model, and the stellar mass and radius are constant during the disc evolution. The stellar mass is selected from a Kroupa IMF distribution (see Sec. \ref{sec:method_models}) and the stellar radius is evaluated from the standard relation $( R_{\star}/\mathrm{R_{\odot}} ) \simeq 1.5 (M_{\star}/\mathrm{M_{\odot}})^{0.6}$,which is valid for young stars. 
    We perturbed the distribution of the resulting accretion luminosities with a spread based on the observed uncertainties in $L_{\mathrm{acc}}$, which are of the order of $0.33$ dex.
    We show the results of the simulations at times included in the age bins of the Orion sub-clusters detailed above, specifically at 1 Myr, 3 Myr, 5 Myr, and 10 Myr, respectively in the panels from left to right.

    In Fig. \ref{fig:histo_fraction}, we compare the detection fraction of $L_{\mathrm{acc}}$ derived from Gaia-based observations with the model predictions. Specifically, the black dots show the percentage of objects with $L_{\mathrm{acc}}$ computed in \citet{Lavinia_Macc} (i.e., the fraction of objects with detected H$_{\alpha}$ in Gaia XP spectra) over the total number of objects presenting Gaia XP spectra, in each FUV flux bin and age bin.
    To estimate the fraction of sources that would be detected in our model based on the observations, we applied to our synthetic accretion luminosities a probability distribution function that, based on the population of stellar luminosities, reproduces the distribution of limit accretion luminosities in the Orion sample. Details of the methodology and the probability function used in the model are provided in Appendix \ref{appendix:lacc_upp}. The results of the models are shown in blue and orange in Fig. \ref{fig:histo_fraction}.\\

    By examining Fig.  \ref{fig:histo_lacc_fuv} and \ref{fig:histo_fraction},  we derive the following main results.
    
    (i) 
    Both model and observations present a distribution of accretion luminosities derived from H$_{\alpha}$ line emission which is shifted towards higher values in younger regions ($<4$ Myr) compared to older regions ($>4$ Myr), as shown in Fig. \ref{fig:histo_lacc_fuv}. 
    The fraction of detected $L_{\mathrm{acc}}$ (Fig. \ref{fig:histo_fraction}) drops more rapidly for stars exposed to stronger FUV fluxes than for those in weaker FUV-irradiated environments.     
    The absence of old sources with detected H$_{\alpha}$ at high FUV fluxes suggests that old YSOs may have dispersed their discs enough for accretion to fall below the H$_{\alpha}$ detection threshold, preventing us from detecting them.
    This may reflect a combined effect of age and strong external irradiation, in the case of similar initial conditions for the population of YSOs across the Orion complex. 

    (ii) The model overestimates the detection fraction in all age bins considered, as shown in Fig. \ref{fig:histo_fraction}.
    This result is particularly interesting because it may indicate that the efficiency of the accretion process of the disc population may lack an important dependence on time in the framework used and may vary as a function of additional external and internal factors (e.g., different disc evolution framework, cluster dynamics).
    However, this is true if we are modelling the same population of accretors and the detection process. We highlight that we modelled a parametric detection curve of accretion luminosities which depends on stellar luminosities, whereas the intrinsic complexity of the Gaia-based detection process and the true underlying completeness may still introduce limitations in the inferred detection fractions. 
    Our investigation highlights the need for more accurate measurements of stellar accretion properties in Orion.
    In Sec. \ref{subsec:discussion_accretion_rate} we further discuss this point.

    (iii) The leftmost panel of Fig. \ref{fig:histo_lacc_fuv} shows that the young stellar population ($<2$ Myr) presents a sample of strong accretors ($L_{\mathrm{acc,median}} > 0.01 \ \mathrm{L}_{\odot}$) that are externally irradiated by strong FUV radiation fields ($>10^{4} \ \mathrm{G}_{0}$). 
    Our models fail to reproduce this young population of strong accretors experiencing high FUV fluxes. Introducing shielding in the first 1 Myr helps to explain the median $L_{\mathrm{acc}}$ in the highest FUV flux bins (see orange squares), but it still cannot perfectly recover the observations, and $L_{\mathrm{acc}}$ remains underestimated. Moreover, a delayed photoevaporation leads to a larger discrepancy between the model and observations for the oldest population ($\gtrsim4$ Myr). We discuss in Sec. \ref{subsec:discussion_accretion_rate}
    that additional factors may act on this young population in a strongly irradiated environment.\\

   The lack of a tight correlation between accretion luminosity and FUV flux in each of the age bins suggests that $L_{\mathrm{acc}}$ is not a preferred tracer of external photoevaporation in the FUV flux range explored (see also Appendix \ref{appendix:lacc_fuv_mr}, Fig. \ref{fig:lacc_fuv_mr}). However, the results shown in Fig. \ref{fig:histo_lacc_fuv} and \ref{fig:histo_fraction} provide valuable insights into the observed sample and the disc evolution framework.
   We provide a detailed discussion of the results in Sec. \ref{subsec:discussion_accretion_rate}. 

\section{Discussion}\label{sec:discussion}
Accurately estimating the FUV flux at the position of stars is fundamental to investigate the connection between disc evolution and the external environment. 
We focused on Orion, the closest star-forming region that hosts a larger number of OBA-type stars and exhibits higher stellar densities compared to more nearby ($<200$ pc) and surveyed regions (e.g. Lupus, Taurus).
We evaluated the FUV flux and its uncertainty at the position of the stars members of the 22 main sub-clusters identified in Orion in a region of a six-degree radius around the ONC core, and we compiled a publicly available table (See Sec. \ref{sec:results_orion_map} and Appendix \ref{appendix:table_fuv}). We discuss the results and the application of the findings in Sec. \ref{subsec:where_to_look_next}, and the limitations and future improvements of the method. 
The relation between stellar accretion and external disc depletion is discussed in Sec. \ref{subsec:discussion_accretion_rate}.
    \subsection{Where to look next?}\label{subsec:where_to_look_next}
    The FUV flux characterization of Orion that we presented in this work is primarily driven by the need to define the range of FUV fluxes covered by Orion, which is fundamental to identify suitable target regions and YSOs to characterise the impact of environmental FUV radiation on disc evolution.  
    We showed that Orion is not the ideal region to target for characterizing the effect of extreme environmental radiation (i.e. $>10^{4} \ \mathrm{G}_{0}$) on disc evolution on a statistical basis, because most of the population is illuminated by moderate FUV radiation fields $\lesssim10^{2} \ \mathrm{G}_{0}$, as shown in Fig. \ref{fig:Orion_map_sub_G0} and \ref{fig:violin}. Although the extended and old population of Orion is subject to FUV fluxes in the range $1-10 \ \mathrm{G}_{0}$, similar to those found in the more nearby star-forming regions (within $\sim200$ pc from the Sun) such as Lupus and Upper Scorpius, we found that $\sim35\%$ of the analysed sample is illuminated by FUV fluxes in the range $10^{2}-10^{4} \ \mathrm{G}_{0}$, which is the typical levels of FUV flux for star-forming regions in the solar neighbourhood (within 2 kpc from the Sun). Therefore, Orion holds potential for further observations particularly aimed at characterising protoplanetary disc evolution and planet formation at FUV radiation fields from low to intermediate levels.
    Candidate targets could be selected by combining our sub-cluster identification and H$_{\alpha}$ emission of \citet{Lavinia_Macc}, which is a tracer of ongoing accretion and, potentially, of a circumstellar disc.
    We comment on some suitable regions for future investigation as follows.
    \begin{itemize}
        \item In Orion A, the extended part of the ONC, outside the Trapezium area, is a suitable target region for studying intermediately FUV-irradiated objects. In this region, the FUV flux is lower than $10^{4} \ \mathrm{G}_{0}$ and proplyds are not detected. For a sample of Class II objects, spectroscopic observations with VLT/X-Shooter provide accurate measurements of stellar and accretion properties \citep{Piscarreta_2025}. Enlarging this spectroscopic sample, as well as future observations aimed to constrain disc properties (e.g., disc dust and gas content, disc extent) are fundamental steps to assess the role of the environment at intermediate FUV radiation fields.   
        \item The South-West part of the ONC includes the L1641 cloud, which is located close to L1647 and is weakly FUV-irradiated ($\sim 5-1000$ G$_{0}$). This part of Orion provides an ideal environment for studying the properties of young discs exposed to FUV fluxes more intense than in Taurus or Lupus, and more similar to those in Upper Sco.
        The Survey of Orion Disks with ALMA (SODA) observed 873 protoplanetary discs in mm-continuum emission in L1641 and L1647 \citep{Soda}, with a detection fraction of $\sim58\%$. They showed a dependence of the median dust disc mass on the FUV flux, despite the weak UV radiation field \citep{Soda_2023}. However, deeper observations and further measurements of disc and stellar properties are needed to assess the role of external photoevaporation on disc dispersal in this region.
        \item In Orion B, NGC 2068 presents a young and unexplored stellar population around the B-type star HD 38563, with FUV fluxes reaching $> 10^{2} \ \mathrm{G}_{0}$.
        \item Orion OB1a contains older clusters ($> 6$ Myr) such as 25 Orionis, and $\psi$ Orionis, and many B-type stars in a spread and low-density region. 
        The fact that the B-type stars in this region are spread across a vast area is a potential consequence of cluster dynamics and expansion across time.
        Exploring YSOs in 25 Orionis and the surrounding clusters, such as Ori CN, would help characterise the advanced stages of moderately irradiated discs.     
    \end{itemize}
    Our investigation suggests that the extended ONC is the best candidate region in Orion to investigate intermediately irradiated objects experiencing $\sim 10^{2} - 10^{4} \ \mathrm{G}_{0}$.  
    Moreover, we emphasise that, to investigate disc evolution across ages, it is crucial to compare young and old disc populations subjected to similar FUV radiation levels. The oldest and closest disc population that has been well-investigated by disc surveys is Upper Sco, which is moderately irradiated, with FUV fluxes in the range $\sim5-100 \ \mathrm{G}_{0}$ \citep{Anania_25}, and hosts numerous B-type stars compared to younger regions such as Lupus and Taurus. In this context, the young and weakly irradiated regions of Orion would provide, to first order, a more appropriate representation of the early evolutionary stages of Upper Sco discs, enabling us to further characterize disc evolution.    
    \subsection{Limitations on the cluster selection and future improvements}\label{subsec:discussion_limitations}
    We computed the FUV flux at each star position using the best estimate 3D separation from massive stars, which is extracted from the posterior distribution in Eq. \eqref{eq:prob_rR_final}. Since this probability involves the use of the 2D geometry of stellar clusters to make arguments about their 3D structure, good knowledge of the membership of a certain cluster is needed. In this work, we used the OPTICS clustering algorithm on a stellar sample extracted from Gaia DR3. The main limitations and future improvements of the method are discussed as follows.

    (i) \textit{Membership of the sub-clusters.} 
    The membership of the Orion sub-clusters identified in this work should not be regarded as complete, since the low-mass population and the stars with very uncertain astrometric parameters in the Orion region may be missing from our selection. 
    However, we discussed in Sec. \ref{sec:sub_clustering} that Gaia DR3 presents a good level of completeness in Orion and that for our purpose, membership completeness is required only to the extent that enables us to reliably determine the shape of the density function of the sub-clusters (i.e., the normalised distribution of the separation between pairs of stars, or from the cluster centre in case of a centrally symmetric cluster). Specifically, to compute the FUV flux, the local density function must be defined at the scales corresponding to the average separation between cluster members and the nearest massive stars. To achieve this, we require at least 5-6 stars per bin separation, considering an average bin widths of fractions of a parsec, which translates into a minimum of 30-40 stars per cluster (i.e., at least $\sim 7-8$ bins per cluster).

    In Appendix \ref{appendix:incomplete_membership}, we used a synthetic stellar cluster to show that by removing a large portion of the total stellar sample, if the final cluster consists of al least 30-40 stars, the shape of the normalised density function can still be determined and thus the final FUV flux is not significantly impacted.

    (iii) \textit{Unidentified sub-clusters}.
    Stellar groups sharing similar astrometric parameters but consisting of less than 30 Gaia DR3 stars cannot be distinguished as separated clusters in our calculation. 
    An example is NGC 2024, where most of the stars in the dense and embedded core of the region are not detected by Gaia due to bad astrometry and the small number of Gaia sources in this region are associated by OPTICS to the close $\sigma$ Ori cluster.
    We verified in \citet{Anania_25} that even considering 2MASS members in the dense core of NGC2024, the shape of the density function cannot be reliably accessed at distances greater than $0.1$ pc from the centre. Two potential solutions to this issue are: considering $\sigma$ Ori and NGC 2024 as part of a single cluster and evaluating the FUV fluxes using 3D separations, being aware of the large uncertainties, or using the 2D projected separation from the closest massive stars, providing estimate of the upper limit FUV flux. Since we show in \citet{Anania_25} that the FUV flux upper limit can significantly differ from the true estimate, the first option is preferred, especially if we need to use the FUV flux estimate in a statistical analysis.

    (iii) \textit{Sub-clusters in the ONC}. 
    Similar to what presented in the previous point, 
    the inner (and denser) ONC core, corresponding to the Trapezium, and small clustered regions such as NGC 1977, and the north part of L1641 with NGC 1981 have astrometry not enough accurate to allow the clustering algorithm to reliably kinematically distinguish them. 
    Considering the ONC as a unique large sub-cluster allows us to compute the FUV flux at stars in its external part, accounting for the 2D geometry of the contained substructures into the profile of the normalised density function. The innermost core can be modelled as a centrally concentric region around $\theta^{1}$C, and isolating the members of NGC1977 we can use a separate density function to compute the FUV flux in this cluster. Considering NGC 1977 as part of a big substructured cluster with the rest of the ONC, or isolating it, results in consistent FUV fluxes for its members.\\
    
    This work does not investigate the $\lambda$ Orionis region, which is located in the north-east part of Orion. $\lambda$ Orionis is an old ($\sim 5$ Myr) cluster, spanning $\sim 10$ degree in diameter on the sky, and where a supernova explosion happened $\sim 1$ Myr ago and has been proposed as a trigger for star formation \citep{lambda_ori_scale, lambda_ori_supernova}. The decrease in mm-continuum disc emission with increasing separation from the most massive star was interpreted as indirect evidence of the effect of UV radiation on discs \citep{Ansdell_lambda_ori}. Due to the peculiarities of the cluster and its large angular extent, we excluded this region from this work. However, a calculation of the FUV flux for the Class IIs in this region can be found in \citet{Anania_25}.
    \subsection{Stellar accretion and disc external truncation}\label{subsec:discussion_accretion_rate}
    We explored the relation between stellar accretion luminosities derived from H$_{\alpha}$ line emission using Gaia XP spectra \citep{Lavinia_Macc} and FUV flux across the Orion region.
    We highlight some important caveats to consider when analysing the results that we presented.

    First, the statistics of the accretion luminosities in Orion are largely dominated by the stars that present Gaia DR3 XP spectra but no H$_{\alpha}$ emission ($\sim 80\%$ of the total population). In Fig. \ref{fig:histo_fraction} we show the fraction of detected $L_{\mathrm{acc}}$ from $H_{\alpha}$ emission related to each age bin and FUV flux bin. 
    Within the sample of stars with no detected H$_{\alpha}$ emission, part of the stellar population may be non accreting, while another fraction may be weakly accreting and its $L_{\mathrm{acc}}$ falls below the detection limit. Since Gaia-based accretion luminosities are not retrieved directly from the H$_{\alpha}$ line flux, but using the H$_{\alpha}$ pseudo-equivalent width, as detailed in \citet{Lavinia_Macc}, distinguishing between a non-accreting and a weakly-accreting star is challenging. Therefore, we modelled a parametric detection curve of accretion luminosities based on the stellar luminosities, and computed representative limit accretion luminosities for the stars with Gaia DR3 XP spectra but no $H_{\alpha}$ emission.
    These values represent heuristic accretion luminosities expected for PMS stars in Orion at a given stellar luminosity $L_{\star}$ (see Sec. \ref{sec:method_models} and Appendix \ref{appendix:lacc_upp}). 
    To ensure a meaningful comparison with the models, we applied to the simulated population the same parametric detection function. 

    Second, although detection of H$_{\alpha}$ line emission retains information of stellar accretion, Gaia-based measurements are based on very low resolution spectra and therefore are affected by large uncertainties. An important source of noise in these measurements is the diffuse H$_{\alpha}$ background emission which, in Orion, may significantly complicate the derivation of stellar accretion luminosities \citep{De_marchi_2010, Beccari_2017}. In particular, it becomes challenging to disentangle the nebular contribution from the intrinsic stellar accretion signature.
    We found no evident dependence of the uncertainties in Gaia-based accretion luminosities on the sky location. Overall, the nature of noise in Gaia XP spectra is complex and cannot be easily quantified.  
    More reliable measurements of stellar accretion can be provided by spectrometric observations (e.g., using instruments such as X-Shooter).
    We found only a marginal positive correlation between the accretion luminosities derived from Gaia and X-Shooter spectra in Orion (Fig. \ref{fig:comp_gaia_xshooter_lacc}), with $\sim 95\%$ of the sources agreeing within their 1$\sigma$ uncertainties. We emphasise that additional spectroscopic measurements of stellar accretion in the Orion region are needed to complement our analysis and robustly assess the effect of environmental radiation on stellar accretion.\\

    We analysed the distribution of Gaia-based accretion luminosities and their detection fraction as a function of age and FUV flux (Fig. \ref{fig:histo_lacc_fuv} and \ref{fig:histo_fraction}, respectively), obtaining three main results that we discuss as follows.

    \subsubsection{Finding evidence of disc external photoevaporation}
    The lack of highly FUV-irradiated sources with detected H$_{\alpha}$ in old regions is shown in Fig. \ref{fig:histo_lacc_fuv} and \ref{fig:histo_fraction} and is broadly consistent with the results of the population synthesis model. 
    The fact that in strongly FUV-irradiated environments we lose strong accretors faster than at weaker FUV fluxes may suggest that external photoevaporation contributes to depleting discs, making $L_{\mathrm{acc}}$ fall below the detection limit and reducing the number of detections. 
    Since a major source of noise in Gaia-based $L{\mathrm{acc}}$ measurements is the nebular background $H_{\alpha}$ emission, we expect that the absence of accreting sources in old regions, which are not associated with dense molecular clouds, is not driven by this observational effect.    
    Nevertheless, this result alone cannot be taken as strong evidence of the effect of external photoevaporation on the Orion protoplanetary disc population.
    A complementary investigation of disc properties (e.g., disc masses, radii) as a function of the FUV radiation field, covering a wide range and FUV fluxes, in the Orion region is needed.  

    Moreover, the population synthesis model presented in Sec. \ref{results:stellar_accretion_vs_fuv} shows that comparing $L_{\mathrm{acc}}$ and FUV flux does not provide strong constraints on the relation between accretion (affecting the inner disc) and depletion (truncating the outer disc).
    Instead, comparing the accretion rate and the photoevaporative mass loss rate may serve this purpose more effectively. 
    Indeed, the full information on the effectiveness of external photoevaporation on discs is not retained in the FUV flux alone, but instead depends also on the stellar mass, disc mass, and disc extent. The photoevaporation rate depends on all these factors and would allow us to better assess the connection between inner and outer disc dynamics.
    However, measuring the disc mass loss rate is extremely challenging. A direct measurement is only possible in proplyds, throughout measurements of the ionization front detectable in optical lines \citep[e.g.,][]{winter_photoev, Aru_proplyds}, and using hydrogen recombination lines \citep{Boyden_2025}. In the future, additional wind tracers will provide a direct way to assess the effect of external photoevaporation (e.g., CI, \citealt{Haworth_owen_CI}) and estimates of the mass loss rate.  
    Indirect measurements involve computing the mass loss rate from models (e.g., interpolating the FRIED grid \citealt{FRIEDv2}) and require knowledge of the stellar mass, disc mass, and disc radius. 
    Observations to assess these properties are challenging to be performed in Orion due to the deep sensitivity and high resolution needed; however they are fundamental to characterise disc evolution in typical star-forming environments.

    \subsubsection{Smaller detection fraction than predicted}
    The population synthesis model overestimates the fraction of detected accretors than is observed, in all age bins. 
    Although this result may suggest that the accretion process efficiency is not entirely captured by a viscous and external photoevaporation model, it is important to underline that the uncertainties in the current dataset prevent us from reliably supporting a deficiency in the model.
 
    We discuss here some of the factors that can influence the accretion process in the model. 
    First, our model assumes a constant level of disc turbulence with stellar mass, time, and across the disc extent by using the parametrisation $\alpha = 10^{-3} (M_{\star}/\mathrm{M}_{\odot})$. The level of turbulence in a protoplanetary disc can instead vary locally and on different timescales due to several factors such as late infalls, close encounters, internal winds, and planet formation \citep[e.g., for a review on disc turbulence see ][]{Rosotti_alpha}. 
    However, it is not obvious that varying the time-evolution of the accretion process can fully solve the tension between models and observations. A small accretion rate allows discs to evolve more slowly and survive longer, thus increasing the detection fraction. In contrast, a higher accretion rate leads to more rapid disc dispersal, resulting in a smaller detection fraction at older age but yet not resolving the discrepancy observed at younger ages.
    Another parameter that may play an important role in determining the efficiency of accretion is the disc magnetisation. Time-variation of this parameter can affect the transport of angular momentum across the disc, which drives accretion \citep[e.g.,][]{AGEPRO_VII_tabone}.
    Moreover, the initial conditions adopted in our model may not be representative of the entire Orion population. Because Orion includes several sub-groups that differ in age, number of members, and OBA stars, the properties of the youngest population may not represent those of the oldest population at its earlier evolutionary stages. 
    As a consequence, a single model is insufficient to capture the diversity in evolution frameworks and the main processes driving disc evolution across different FUV radiation fields. 
    Recent works studied the evolution of single protoplanetary discs within an MHD wind-driven framework where external photoevaporation is included \citep{Coleman_2024, Pichierri_2026}, showing that disc dispersal timescale and the relation between stellar accretion rate and disc mass loss rate can be significantly influenced by the wind strength.
    Future work investigating the impact of additional processes such as internal photoevaporation \citep[e.g.,][]{Malanga_2025}, and using an MHD wind-driven framework instead of a viscous evolution scenario \citep[e.g.,][]{AGEPRO_VII_tabone} in a disc population synthesis model will provide further constraints on the preferred model and initial parameter space describing the Orion sub-cluster populations. 

    In addition to the model assumptions that may influence the results, it is essential to emphasise that, since the noise in the Gaia-based accretion luminosity cannot be reliably quantified, our ability to assess the role of external radiation fields on stellar accretion in Orion is significantly limited.

    \subsubsection{Young population of highly irradiated strong accretors}
     We detected a sample of accretors that are young ($\lesssim1-2$ Myr), highly irradiated ($\gtrsim 10^{4} \ \mathrm{G}_{0}$), and have high accretion luminosities ($\gtrsim  10^{-2} \ \mathrm{L}_{\odot}$), which cannot be reproduced by a population model of viscous externally photoevaporating discs. 
    A potential explanation may involve the role of dust extinction in efficiently shielding protoplanetary discs. We showed that a population synthesis model in which the effect of external photoevaporation is delayed by 1 Myr supports in part this hypothesis, although it does not entirely solve the discrepancy because $L_{\mathrm{acc}}$ remains underestimated.
    To address this discrepancy, we discuss scenarios that depend on both model and observations.
    
    On the model-dependent side, we highlight three scenarios. The first involves different initial conditions for our population synthesis model, implying that the parameter space that we explored is not fully representative of both the younger and older Orion population. As mentioned in the previous Section, and recently supported by \citep{Weder_cygnus}, a single disc evolution model faces challenges in explaining properties in different star-forming regions.  
    The second scenario involves the role of cluster dynamics in varying the FUV flux across time. Substantial temporal fluctuations in FUV flux can lead to significant variations in the external photoevaporation mass loss rate and, since the amount of material in the disc varies, it is not excluded that this will affect the accretion process. Considering, for example, an OB star that has been closer or farther by 1 pc from an accreting source at some point in the past, the corresponding FUV flux at the accreting source would have varied by a factor of four. However, since the models do not predict a strong dependence of $L_{\mathrm{acc}}$ on the FUV flux (in the wide range explored, $\sim1-10^{5} \ \mathrm{G}_{0}$), such variation is unlikely to be the dominant factor required to explain the observed accretion luminosities in Orion.   
    The last scenario involves infalling material from the external environment that is able to raise the rate of stellar accretion. 
    To support this last scenario, we notice that the sample of young and highly irradiated sources is located in the ONC core, which is the part of Orion presenting the highest stellar and gas density. The correlation between accretion rate and stellar and gas density was investigated in the Lupus region, revealing that clustered sources present higher accretion rates than those that are non-clustered, and that accretion rates are consistent with the predictions of a Bondi-Hoyle-Lyttleton accretion model \citep{Winter_lupus}. 
    A similar effect may complement dust extinction shielding in explaining the high accretion luminosities detected for the strongest irradiated sources.

    On the observational side, ambient background H$_{\alpha}$ may significantly contribute to the $L_{\mathrm{acc}}$ evaluated from Gaia observations. Since the young sources exposed to strong FUV radiation fields are mainly located in the ONC, where substantial H$_{\alpha}$ is also emitted by the surrounding dense cloud, it cannot be excluded that high L$_{\mathrm{acc}}$ values are at least partly influenced by this background contribution.
\section{Conclusions}\label{conclusions}
Connecting the evolution of YSOs and their protoplanetary discs to the stellar formation environment is essential to characterise the resulting planetary systems.
In this work, we aimed to determining the FUV flux range that is probed by the Orion stellar population, and investigating the relation between stellar accretion and FUV radiation field across the region.

We selected from Gaia DR3 the stars in a region of radius of six degree around $\theta^{1}$C. Using the clustering algorithm OPTICS, we isolated 22 main sub-clusters among the selected stellar population (Fig. \ref{fig:Orion_map_sub_G0}).
We used the 2D geometry of the sub-clusters to evaluate the best estimate 3D separation from OBA-type stars, and then compute the FUV flux and its uncertainty at the stars members of the sub-clusters (see Data availability Section).

We found that 1200 objects in the investigated Orion population have accretion luminosity evaluated from H$_{\alpha}$ line emission using Gaia XP spectra \citep{Lavinia_Macc}. We studied these accretion luminosities as a function of age and FUV flux, and we compared the results with a population synthesis model of viscous discs experiencing external photoevaporation.   
 
The main results obtained are summarised as follows.
\begin{list}{$\bullet$}{}
    %
    \item Most of the Orion stellar population analysed is weakly irradiated, $\lesssim10^{2} \ \mathrm{G}_{0}$ (Fig. \ref{fig:Orion_map_sub_G0}). In particular, stars in Orion OB1a show FUV radiation levels comparable to those in more nearby regions such as Lupus and Taurus, but they are on average older ($> 6$ Myr). 25 Orionis presents a similar age and FUV radiation field of Upper Sco. Future comparison between disc populations experiencing similar FUV flux levels in Orion and in more nearby regions may enhance our understanding of disc evolution across ages and different star-forming environments. 
    \item Orion is the closest region that, while containing numerous weakly irradiated stars, also hosts a substantial subset of intermediately irradiated stars exposed to FUV fluxes in the range $10^{2} - 10^{4} \ \mathrm{G}_{0}$. This intermediately irradiated population accounts for $\sim35\%$ of our sample. In contrast, only $\sim5\%$ of the investigated stellar population is exposed to strong FUV radiation fields ($>10^{4} \ \mathrm{G}_{0}$). To investigate the impact of extreme environments on disc evolution, it is necessary to extend such studies to more distant and massive star-forming regions than Orion.
    \item $L_{\mathrm{acc}}$ from Gaia-based observations decreases with age in Orion (Fig. \ref{fig:histo_lacc_fuv}). The fraction of sources with detected H$_{\alpha}$ line emission declines more rapidly in strongly FUV-irradiated environments ($>10^{2} \ \mathrm{G}_{0}$) than in regions exposed to weaker FUV fluxes ($<10^{2} \ \mathrm{G}_{0}$). This trend is broadly reproduced by the population synthesis model (Fig. \ref{fig:histo_lacc_fuv} and \ref{fig:histo_fraction}) and may be an hint of the effect of external photoevaporation depleting efficiently discs exposed to strong FUV fluxes. 
    \item The population synthesis model overestimates the detection fraction of $L_{\mathrm{acc}}$ with respect to observations, across all the age and FUV flux bins explored. Moreover, the model is unable to fully explain the young population ($<2$ Myr) of strong accretors ($L_{\mathrm{acc}} > 10^{-2} \ \mathrm{L}_{\odot}$) that are highly FUV-irradiated ($>10^{4} \ \mathrm{G}_{0}$), even when the effect of dust extinction in shielding the discs is included.
    We discuss in Sec. \ref{subsec:discussion_accretion_rate} that, although a single model may be insufficient to capture the diversity in disc evolution across environments with weak to strong FUV radiation fields, the main challenge arises from quantifying completeness and noise in the Gaia DR3 XP dataset. Therefore, we highlight that firm conclusions about the relation between stellar accretion and FUV radiation field in Orion require spectroscopic observations of stellar and accretion parameters, supported by measurements of disc properties.    
\end{list}
This work provides new tools and encourages further investigation of stellar and disc properties in the Orion region with the aim of characterising the role of the environment in disc evolution and planet formation across FUV radiation fields.

\section*{Data availability}
Table \ref{appendix:table_table_fuv} is only available in electronic form at the CDS via anonymous ftp to cdsarc.u-strasbg.fr (130.79.128.5) or via \href{http://cdsweb.u-strasbg.fr/cgi-bin/qcat?J/A+A/}{$http://cdsweb.u-strasbg.fr/cgi-bin/qcat?J/A+A/$}.\\
An accessible \texttt{Python} routine allowing the computation of the FUV flux for stars in Orion sub-clusters is available at \href{https://github.com/Rossella4712/Orion_FUV}{$https://github.com/Rossella4712/Orion_FUV$}. 

\begin{acknowledgements}
RA and GR acknowledge funding from the Fondazione Cariplo, grant no. 2022-1217, and the European Research Council (ERC) under the European Union’s Horizon Europe Research \& Innovation Programme under grant agreement no. 101039651 (DiscEvol). Views and opinions expressed are however those of the author(s) only, and do not necessarily reflect those of the European Union or the European Research Council Executive Agency. Neither the European Union nor the granting authority can be held responsible for them.
GL acknowledges support from PRIN-MUR 20228JPA3A and from the European Union Next Generation EU, CUP: G53D23000870006

\end{acknowledgements}
\bibliographystyle{bibtex/aa} 
\bibliography{bibtex/mybib}

%





\begin{appendix}

\section{Effect of an incomplete cluster membership on the FUV flux calculation}\label{appendix:incomplete_membership}
\begin{figure}
        \centering
        \includegraphics[width=0.3\textwidth]{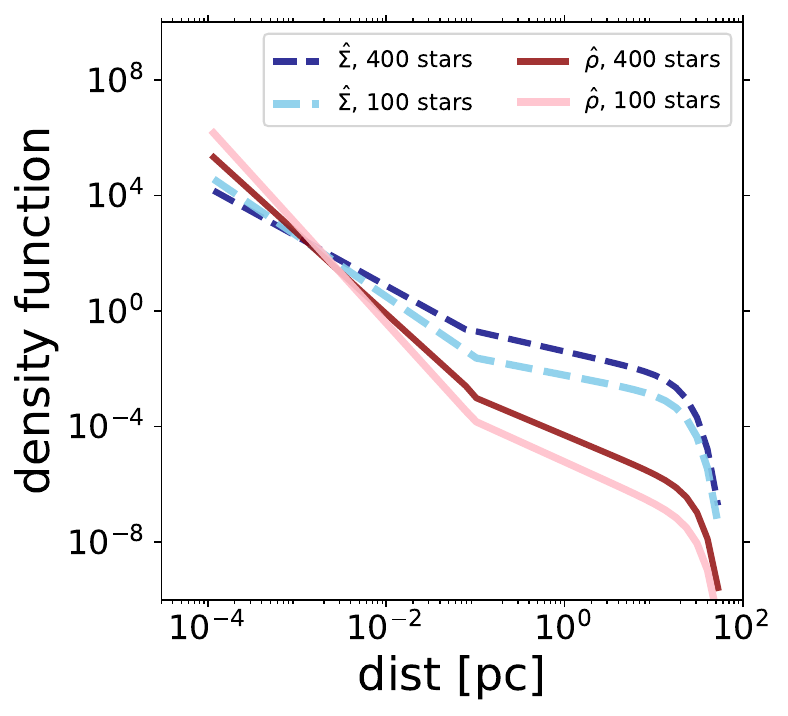}
        \hspace{0.4cm}
        \includegraphics[width=0.28\textwidth]{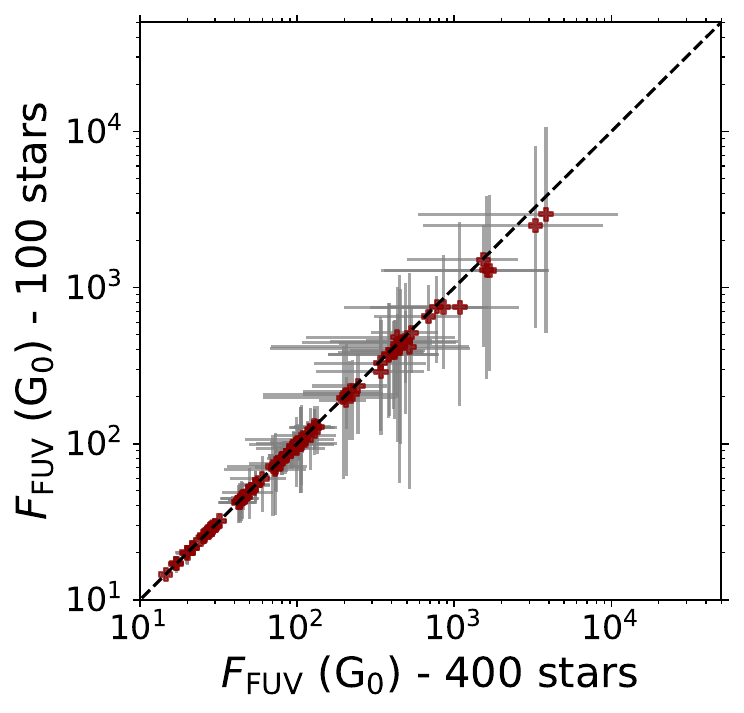}
        \caption{Left: Normalised surface (dashed lines) and volume (solid lines) local density function of a synthetic cluster hosting 400 stars (dark blue and red), and when removing 300 stars randomly (light blue and pink). Right: Relation between the FUV fluxes (and the uncertainties, shown as the 16$^{\mathrm{th}}$ and the 84$^{\mathrm{th}}$ percentiles) computed using the local density function of a cluster with 400 stars (x-axis) and with 100 stars (y-axis). The dashed lines indicate a perfect correlation (1:1).  }
        \label{fig:reduced_membership}
    \end{figure}

    We used a synthetic stellar cluster to investigate the effect of membership incompleteness on the resulting FUV flux. We considered a substructured synthetic cluster that initially contains 400 stars (built as detailed in \citealt{Anania_25}). Then, we removed randomly 300 stars, without impacting the OBA stars that provide the irradiation, and computed the local density function and the FUV flux at the position of the remnant stars. In the left panel of Fig. \ref{fig:reduced_membership}, we show that the local density function profile is not significantly impacted by the removal of members. Consequently, we obtain a similar FUV flux, as shown in the right panel of Fig. \ref{fig:reduced_membership}. However, with a sample of $\lesssim 40$ stars, deriving the shape of the density function is challenging and the final result may be not accurate, leading to FUV fluxes affected by large uncertainties.
    Incomplete membership may have a stronger impact on smaller separations than on larger ones, particularly in areas with high extinction. In our calculation, we required at least 5-6 stars per bin separation, with average bin width of fractions of a parsec, and we further ensured that the smallest separation bin adequately captures the determination of the power-law slope at small separations, as shown in Fig. \ref{fig:reduced_membership}. 
    With this approach, highly extincted and strongly incomplete sub-clusters that cannot be reliably isolated are incorporated into larger sub-cluster structures. The resulting memberships and density profiles can be refined in the future as more complete clustering information becomes available.
    
\section{Table of the FUV flux on stellar members of Orion}\label{appendix:table_fuv}
We provide Table \ref{appendix:table_table_fuv} containing the FUV fluxes at the position of the stars members of the Orion sub-clusters identified in this work. The full version of the table is available at the CDS.

\begin{sidewaystable*}[]
\caption{Table of the FUV fluxes for the Orion members of the 22 sub-clusters}\label{appendix:table_table_fuv}
\begin{tabular}{c |c |c |c |c |c |c |c |c |c |c |c}
\hline \hline
     Gaia DR3  & cluster name & RA & Dec  & pmRA & pmDec & d  & d 16 & d 84 & FUV flux & FUV flux 16 & FUV flux 84 \\
      & & (deg) & (deg) & (mas yr$^{-1}$) & (mas yr$^{-1}$) & (pc) & (pc) & (pc) & (G$_{0}$) &  (G$_{0}$) &  (G$_{0}$) \\
\hline
3012327436772983936 & eta\_Ori      & 85.93179 & -8.57125  & 0.42755         & -0.86023         & 420.81885 & 411.82153    & 430.6435     & 3.34538       & 2.78001           & 4.03291           \\
3012331598597029632 & eta\_Ori      & 85.71879 & -8.65466  & 0.36027         & -0.93494         & 422.28973 & 411.52426    & 435.362      & 3.33534       & 2.72817           & 4.05422           \\
3015418099535217024 & eta\_Ori      & 85.37746 & -8.08008  & 0.31999         & -1.1983          & 416.0579  & 403.87207    & 427.73135    & 6.7279        & 4.5098            & 8.61105           \\
3015729050872503296 & eta\_Ori      & 85.69316 & -8.20644  & -0.07414        & -1.25782         & 419.80786 & 396.35812    & 456.41824    & 7.57145       & 4.26048           & 11.39487          \\
3015742902142721280 & eta\_Ori      & 85.70453 & -8.11751  & 0.47192         & -1.09291         & 413.58554 & 410.48248    & 417.6859     & 21.01749      & 5.79932           & 41.51186          \\
3015732108888685824 & eta\_Ori      & 85.96573 & -8.18792  & 0.44456         & 0.12445          & 404.26547 & 396.694      & 414.59558    & 5.01758       & 4.06586           & 6.00264           \\
3015732869099186048 & eta\_Ori      & 85.91195 & -8.1781   & 0.51854         & 0.07494          & 404.18936 & 395.41833    & 411.2511     & 5.47594       & 4.27296           & 6.62207           \\
3015786539010363520 & eta\_Ori      & 86.13789 & -7.6568   & 0.29596         & 0.13072          & 405.32208 & 400.863      & 409.29132    & 4.98657       & 4.1296            & 5.85834           \\
3012342181396446080 & eta\_Ori      & 85.62757 & -8.54869  & 0.53739         & -0.17075         & 414.98343 & 412.99976    & 417.4557     & 3.88659       & 3.288             & 4.59409           \\
3015352055822114432 & eta\_Ori      & 85.66883 & -8.30432  & -0.12019        & -0.81861         & 403.99768 & 374.6989     & 434.98822    & 5.45121       & 4.02657           & 6.8451            \\
3015740733183554560 & eta\_Ori      & 85.62863 & -8.17933  & -0.02527        & -0.24276         & 404.61987 & 385.9944     & 428.1689     & 16.13003      & 5.95683           & 32.65284          \\
3015741562113224704 & oriC\_01      & 85.64549 & -8.13069  & 0.24641         & -0.56287         & 407.84277 & 382.36554    & 434.93082    & 42.80094      & 7.24743           & 124.44825         \\
3015826705544481152 & oriC\_01      & 85.44502 & -7.57173  & 0.25675         & -0.40004         & 407.75348 & 400.53107    & 414.32132    & 5.15001       & 4.06775           & 6.39686           \\
3015741424673978624 & oriC\_01      & 85.62832 & -8.15154  & -0.05306        & -0.39531         & 435.7206  & 419.01996    & 454.3805     & 34.37116      & 6.18125           & 81.97744          \\
3012343792008175360 & oriC\_01      & 85.84826 & -8.44456  & 0.36365         & -0.82941         & 410.68442 & 384.30783    & 437.8266     & 3.83173       & 2.93141           & 4.6712            \\
3012704676635505792 & oriC\_01      & 86.16384 & -8.44442  & 0.57235         & -0.22403         & 403.4936  & 385.88632    & 421.31363    & 3.65097       & 2.98841           & 4.30477          \\
3012722509340910336 & oriC\_01      & 85.94376 & -8.31449  & 0.34329         & -0.73505         & 397.5587  & 390.89233    & 405.26056    & 4.43917       & 3.73325           & 5.18455           \\
3012654992455074176 & oriC\_01      & 86.45786 & -8.5862   & 0.48702         & -0.08344         & 401.9201  & 395.7014     & 412.17633    & 3.3388        & 2.84897           & 3.89879           \\
3012700892769837952 & oriC\_01      & 86.21824 & -8.52018  & 0.56253         & 0.11573          & 395.78912 & 382.81808    & 407.4445     & 3.61914       & 3.04598           & 4.23606           \\
3012222815665441024 & oriC\_01      & 85.66596 & -9.36283  & 0.26643         & -0.70582         & 428.12418 & 420.45706    & 436.0316     & 2.66956       & 2.23223           & 3.15625           \\
3012331147624652032 & oriC\_01      & 85.67414 & -8.68456  & 0.23998         & -0.91891         & 456.1228  & 389.4239     & 528.87476    & 2.26738       & 1.01889           & 3.79658          \\
3015421157551880064 & oriC\_01      & 85.15566 & -8.06751  & 0.01801         & -1.229           & 429.69632 & 426.941      & 431.99667    & 4.60933       & 3.65971           & 5.61271           \\
3015822616735611008 & oriC\_01      & 85.57678 & -7.62983  & 0.08493         & -0.99752         & 451.24313 & 392.8971     & 526.15375    & 2.99971       & 1.21454           & 5.12012           \\
3015758673262683392 & oriC\_01      & 86.1943  & -8.02415  & 0.55793         & -0.56535         & 395.08273 & 386.8568     & 405.7362     & 4.98654       & 4.15943           & 5.86122           \\
3222162191483504512 & psi\_Ori      & 81.06736 & 1.51621   & -0.28204        & 1.3137           & 420.59763 & 415.1086     & 426.24155    & 2.2955        & 2.01902           & 2.63198           \\
3222180814461726976 & psi\_Ori      & 80.76958 & 1.62081   & -0.1926         & 0.90442          & 415.29898 & 405.18735    & 425.03146    & 2.31156       & 1.96456           & 2.7803            \\
3222218983835355904 & psi\_Ori      & 81.51833 & 1.70659   & -0.0269         & 0.45164          & 448.0136  & 404.6206     & 526.1942     & 1.67565       & 0.8696            & 3.28599           \\
3222223832854640768 & psi\_Ori      & 82.00571 & 1.65285   & -0.18482        & 0.83404          & 410.732   & 400.9983     & 422.16293    & 3.02608       & 2.54399           & 3.77162           \\
3221083330058766208 & psi\_Ori      & 82.81532 & 0.34053   & 0.18486         & 1.69534          & 396.0425  & 383.62518    & 406.19635    & 5.1472        & 4.40311           & 5.92679           \\
3223741979240153856 & psi\_Ori      & 82.41087 & 1.96659   & -0.26615        & 1.16389          & 411.5856  & 392.21173    & 436.9492     & 2.87246       & 2.07462           & 3.8839            \\
... &... &... &... &... &... &... &... &... &... &... &...\\

\hline
\end{tabular}%
\tablefoot{\footnotesize{The columns of the table indicates, for the objects members of the 22 Orion sub-clusters identified in this work (Sed. \ref{sec:sub_clustering}, Fig. \ref{fig:Orion_map_sub_G0}), from left to right, Gaia DR3 identificator, name of the sub-cluster, RA, Dec, $\mu_{\mathrm{RA}}$, $\mu_{\mathrm{Dec}}$, Bailer-Jones distance (median, 16$^{\mathrm{th}}$ and 84$^{\mathrm{th}}$ percentiles), and FUV flux best estimate (median, 16$^{\mathrm{th}}$ and 84$^{\mathrm{th}}$ percentiles). The extended version of the table can be found at the CDS.}}
\end{sidewaystable*}

\section{Deriving the FUV flux from dust emission}\label{appendix:sut_comparison_maps}
    Another method for computing the FUV flux in young star-forming regions has been previously introduced by \citet{Hollenbach_g0_comp} and consists of linking the dust temperature with the FUV radiation field. \citet{Hollenbach_g0_comp} used a theoretical 1D slab model that solves thermal and chemical balance considering incident FUV radiation on one side, to derive the relation between the dust temperature and the FUV flux in units of G$_{0}$:
    \begin{multline}
        T_{\mathrm{dust}} = \{ 8.9\times10^{-11} \ \nu_{0} \ \mathrm{G}_{0} \ e^{-1.8A_{v}} + 2.7^{5} + 3.4\times10^{-2} \\
        [0.42 - \ln{(3.5\times10^{-2} \tau_{100}T_{0})}]
        \times\tau_{100} T_{0}^{6} \}^{0.2},
        \label{eq:Tdust_g0}
    \end{multline} 
    where $\nu_{0} = 3\times10^{15}$ Hz, $T_{0}$ is the equilibrium dust temperature at the slab surface due to the incident FUV photons, $T_{0} = 12.2\ \mathrm{G}_{0}^{0.2}$, and $\tau_{100}=(2.7\times10^{2}\ \mathrm{G}_{0})/T_{0}^{5}$ is the effective optical depth at 100$\mu$m that, accordingly to the model of \citet{Hollenbach_g0_comp}, takes into account the attenuation of the FUV photons through the dust grains.
    The \textit{Herschel} Gould Belt survey (HGBS, \citeauthor{Andrè_HGBS}\citeyear{Andrè_HGBS}; \citeauthor{Konyves_HGBS}\citeyear{Konyves_HGBS}) released the dust temperature maps for Orion A and B \citep{Pezzuto_tdust}\footnote{\url{http://www.herschel.fr/cea/gouldbelt/en/Phocea/Vie_des_labos/Ast/ast_visu.php?id_ast=66}}, that we used to compute G$_{0}$ from the expression in Eq. \eqref{eq:Tdust_g0}. \\
    
    There is also an alternative method involving the use of dust emission for computing G$_{0}$, which was proposed by \citet{Kramer_g0_calc} and is based on the assumption that the FUV radiation is completely absorbed by interstellar dust grains and re-emitted at FIR wavelengths. The FIR intensity, which will be converted into G$_{0}$, can be computed using the FIR maps at 70$\mu$m and 160$\mu$m provided by the HGBS \citep{Ishii, Shimajiri_HGBS_g0_calc}. However, since the Herschel broadband filters do not provide full coverage of the entire FIR, a significant part of the radiation may be missing from the calculation, resulting in significantly underestimated FUV fluxes. Therefore, we did not focus on this method, while we did compare our proposed method with the approach described by Eq. \ref{eq:Tdust_g0}.\\

\begin{figure}
    \centering
    \includegraphics[width=0.50\textwidth]{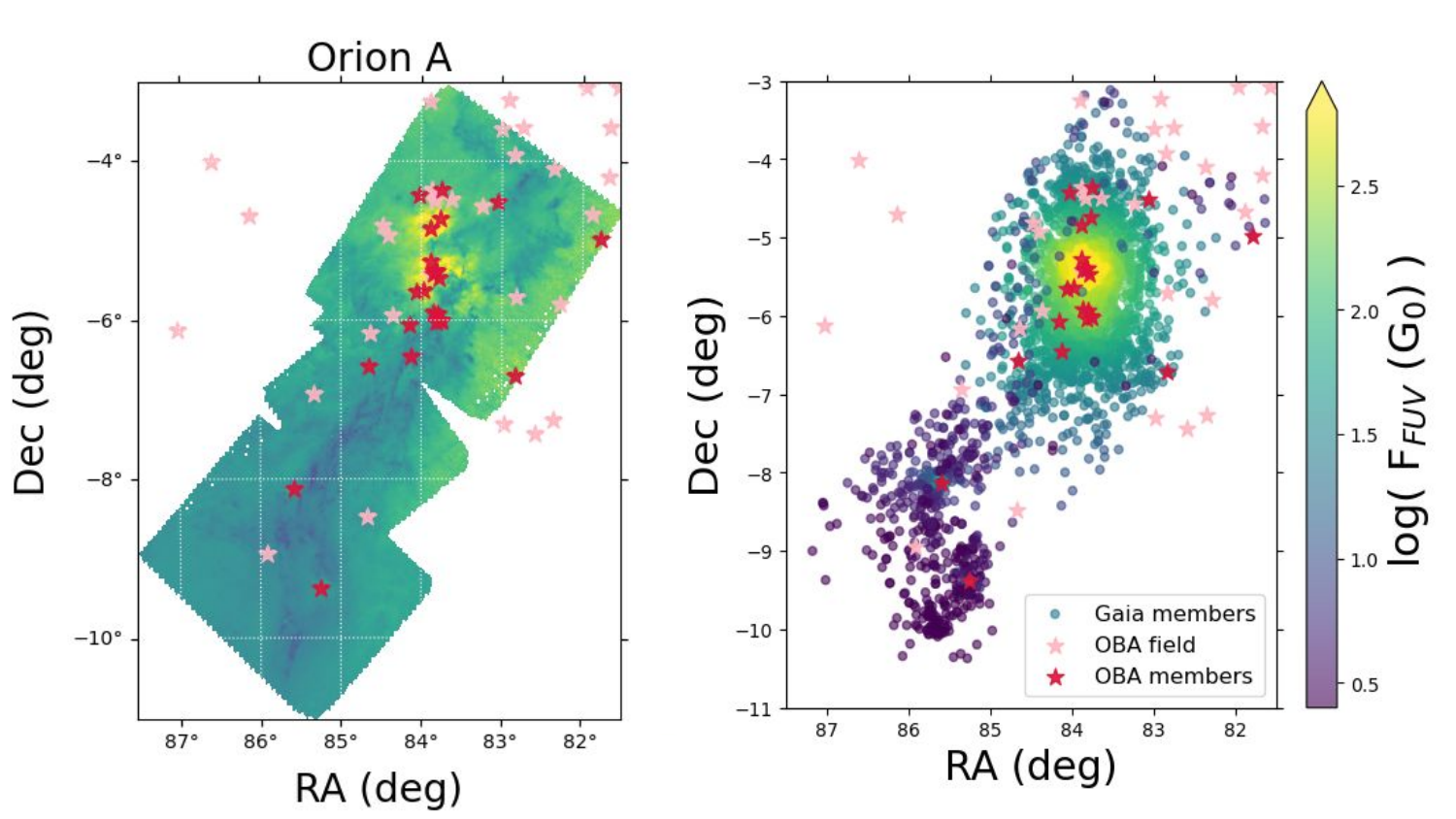}
    \includegraphics[width=0.50\textwidth]{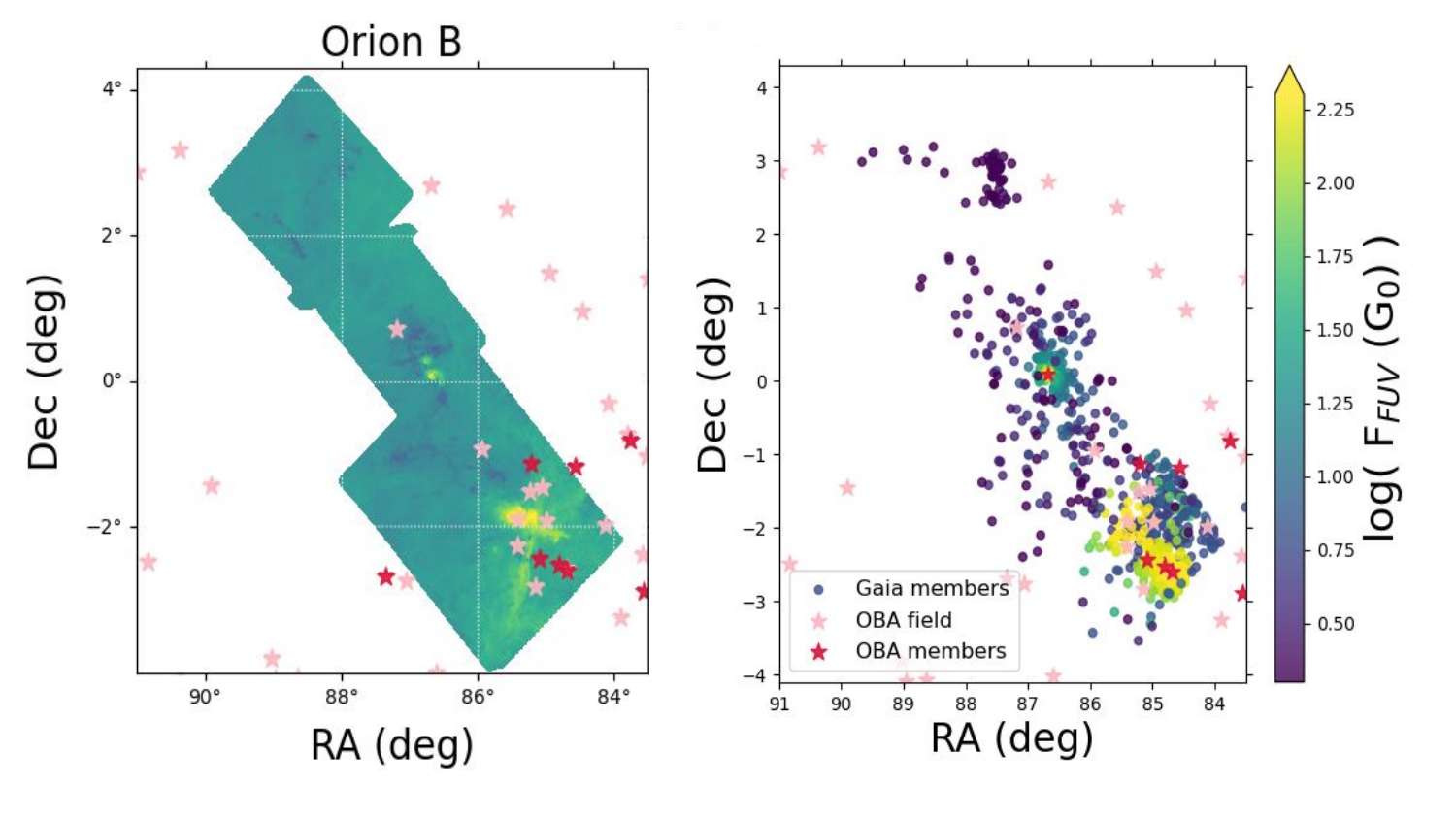}
    \caption{Top row: FUV flux map of Orion A as derived from the HGBS map of the dust temperature using Eq. \eqref{eq:Tdust_g0} (left panels), and FUV flux at the positions of the stars included in that area, using the local density function method described in Sec. \ref{subsec:FUV_flux_comp_density} (right panels). OBA-type stars are marked in red (members) and pink (field stars). Bottom row: Same as top row but for Orion B.}
    \label{fig:Orion_A_B_HGBS}
    \end{figure}

    Herschel FIR observations do not cover the entire Orion region explored in this study. However, since the HGBS mapped Orion A and Orion B, providing the column density and dust temperature maps, we computed the FUV flux for these regions.
    In the panels of the left column of Fig. \ref{fig:Orion_A_B_HGBS}, we show the FUV flux maps of Orion A and Orion B resulting from using the relation between the dust temperature provided by the HGBS and G$_{0}$ (Eq. \ref{eq:Tdust_g0}). The HGBS dust temperature maps cover Orion A and B with a resolution of 36.3$''$. In the panels of the right column of Fig. \ref{fig:Orion_A_B_HGBS}, we present the FUV flux at each stellar position covered by the Orion A and Orion B dust maps, computed applying the local density function method. 

\begin{figure}
    \centering
    \includegraphics[width=0.35\textwidth]{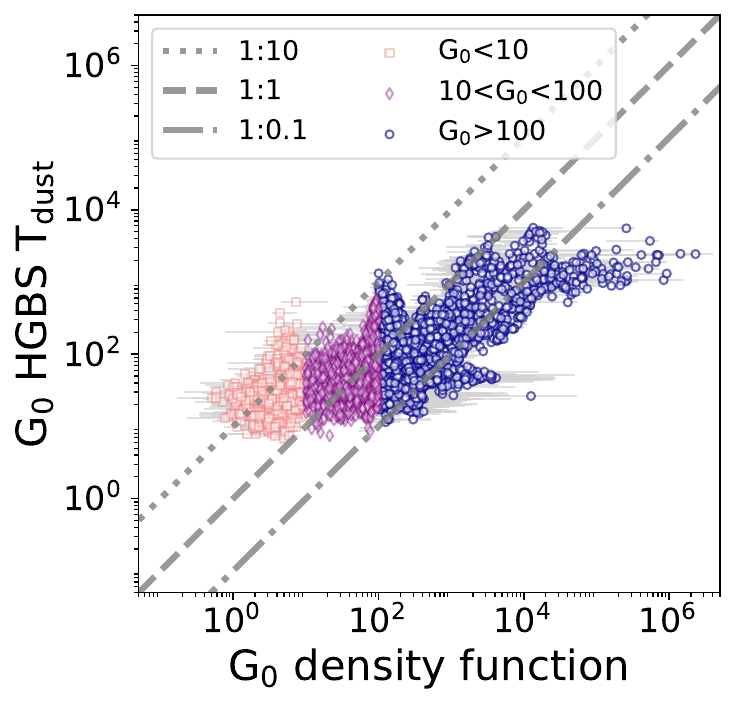}
    \caption{Comparison between the FUV flux values (in units of G$_{0}$) computed using the HGBS dust temperature maps of Orion A and Orion B, using Eq. \eqref{eq:Tdust_g0}, and those resulting from applying the local density function method (Sec. \ref{subsec:FUV_flux_comp_density}). Gray lines indicates where G$_{0}$ computed from dust temperature is [10, 1, 0.1] times the G$_{0}$ resulting from local density function. Values provided by using dust temperature maps should be more reliable for higher G$_{0}$.}
    \label{fig:comp_G0_Tdust_HGBS_density}
    \end{figure}

    We interpolated the FUV flux map derived from Herschel at the position of our stellar sample to make a comparison between the results of the two methods. The comparison is shown in Fig. \ref{fig:comp_G0_Tdust_HGBS_density}\footnote{The comparison is made byinterpolating the 2D dust temperature map at the star positions where we evaluated the FUV flux. However, our FUV flux calculation considers the 3D geometry, and therefore the comparison may be influenced by projection effects}. We found that the FUV flux values derived from dust temperature are consistent within 1$\sigma$ error with those evaluated by applying the local density function method. The $\sim 30\%$ of the sample show consistent FUV fluxes within a factor of 2, while the remnant sample tends, in general, to predict a larger G$_{0}$, remaining within an order of magnitude error. The main limitations in computing G$_{0}$ using dust emission maps are summarised as follows:

    (i) G$_{0}$ resulting from dust maps depends on the amount of dust in the region. In areas with very low or no dust, the computed FUV flux is artificially low (i.e. not because the radiation is weak or there are no massive stars in the region, but because there is insufficient dust to trace it). As a result, low G$_{0}$ can be found near massive stars if they lie in dust-poor regions (e.g. very old regions). This is evident in the comparison made for Orion B, in Fig. \ref{fig:Orion_A_B_HGBS}, where the left plot shows low G$_{0}$ values in correspondence with the $\sigma$ Orionis cluster, where only small amounts of dust are present, probably due to the older age ($\sim3$ Myr) with respect to NGC 2024 and the Orion A region ($<2$ Myr). G$_{0}$ is predicted to be higher in NGC 2024, where the dust density is very high. In this last cluster, we evaluated G$_{0}$ only for a few Gaia stars which are not close enough to the most massive star to reach $>10^{4} \ \mathrm{G}_{0}$, which is instead registered using the dust temperature map.
    
    (ii) As discussed in \citet{Hollenbach_g0_comp}, below $\sim 10 \ \mathrm{G}_{0}$ the dust is mainly heated by visible and far-infrared photons, whereas FUV photons do not contribute to the majority of the emitted FIR radiation and the measured dust temperature. 
    At higher, but moderate, FUV flux level between 10 and 100 G$_{0}$, the dust heating can still be largely contaminated by visible and far-infrared. Therefore, using HGBS maps in moderate-to-low G$_{0}$ regions may lead to an overestimate of the actual FUV flux value. In Fig. \ref{fig:comp_G0_Tdust_HGBS_density}, we marked in pink and purple the points at $F_{\mathrm{FUV}} <10\ \mathrm{G}_{0}$ and $10 \ \mathrm{G}_{0}< F_{\mathrm{FUV}} < 100\ \mathrm{G}_{0}$, corresponding to low and moderate probability of accurately describing the actual FUV flux, while colouring in blue the region where the FUV flux estimate obtained by using Herschel maps is more reliable.   

    (iii) When using the relation between FUV flux and dust temperature described in Eq. \ref{eq:Tdust_g0}, the main source of uncertainty is the extinction term $A_{\mathrm{v}}$, which is usually associated with large uncertainty in young and embedded regions.

    (iv) The relation between dust temperature and G$_{0}$ shown in Eq. \eqref{eq:Tdust_g0} is based on 1D slab models, where UV radiation is incident on one side of the slab \citep{Hollenbach_g0_comp}. 
    In the worst-case scenario, not including dust extinction when computing the FUV flux from the local density function can result in an order-of-magnitude difference in the FUV flux. However, in less extreme (and more probable) cases, ignoring the influence of 3D geometry and thus using a 1D slab model may contribute to a general trend for our method in predicting higher FUV fluxes, as shown by the blue region in Fig. \ref{fig:comp_G0_Tdust_HGBS_density}.\\

    Overall, although using dust emission maps to evaluate G$_{0}$ gives an indicative value of the average FUV radiation field in the region of interest, it has numerous limitations, from the number of regions in which it can be applied, which is limited to the regions mapped in dust emission and FIR with good resolution, to the application to older regions, where dust density is low and the result is not reliable, to the model-dependent limitations discussed in the list above. Conversely, our method relies on considerations on the 2D geometry of a stellar cluster. This can lead to intrinsic large error bars in the FUV flux depending on the shape of the local density function: the flatter the local density function is, the flatter the associated cumulative distribution function (CDF) will be, and therefore, larger will be the range of most probable 3D separations at which two stars can be.
    However, our method provides a way to consider the main source of uncertainty in the FUV flux calculation, which is the 3D structure of the stellar cluster, providing an estimate of the uncertainty in the FUV flux. The FUV flux distribution at the position of each star can be used in statistical analysis to investigate, more accurately, the connection between protoplanetary disc properties and the external environment. 


\section{Comparing Gaia and X-Shooter based accretion luminosities}\label{appendix:compare_gaia_xshooter_lacc}
\begin{figure}
        \centering
        \includegraphics[width=0.33\textwidth]{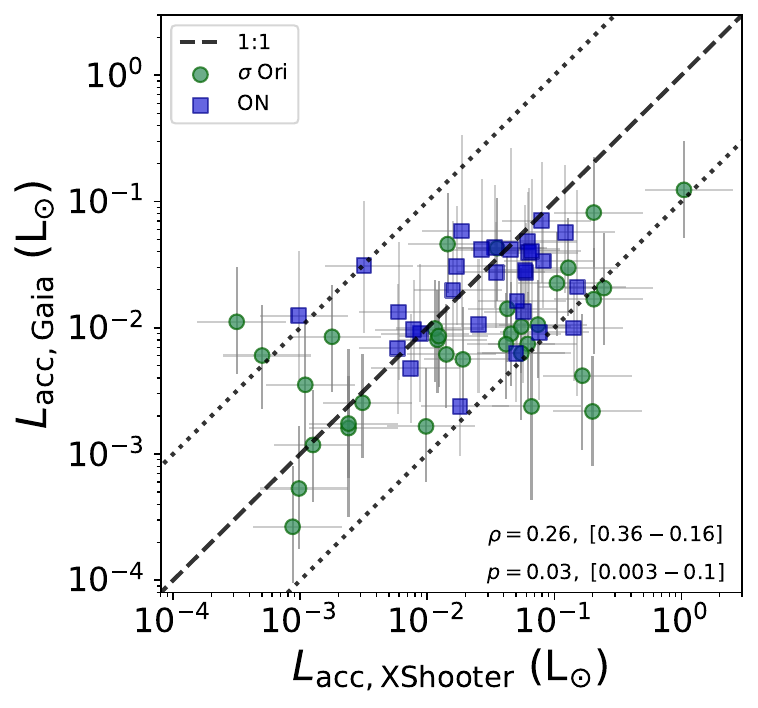}
        \caption{Comparison between accretion luminosities measured from X-Shooter spectra and those measured from $H_{\alpha}$ emission from Gaia XP spectra \citep{Lavinia_Macc} in the $\sigma$ Orionis \citep{Mauco_sigmaOri} and the Orion Nebula (ON, \citealt{Piscarreta_2025}) regions. The dashed line marks the 1:1 relation, while the dotted lines indicate 1$\sigma$ uncertainty from the 1:1 line. The $\rho$ and p-value Spearman coefficients of correlation are shown in the bottom right, with the $16^{\mathrm{th}}$ and $84^{\mathrm{th}}$ percentile interval.
        }
        \label{fig:comp_gaia_xshooter_lacc}
    \end{figure} 
    Figure \ref{fig:comp_gaia_xshooter_lacc} presents the comparison between the accretion luminosities of the sample of objects which have X-Shooter spectra in Orion (i.e., in the $\sigma$ Ori,  \citealt{Mauco_sigmaOri}, and ON \citealt{Piscarreta_2025}, regions) and its counterpart in \citet{Lavinia_Macc}.
    The Spearman rank correlation coefficient $\rho =0.26$ ($16^{\mathrm{th}} -84^{\mathrm{th}} \ \mathrm{percentile \ range} = 0.36-0.16$), and the p-value $p = 0.03$ ($16^{\mathrm{th}} -84^{\mathrm{th}} \ \mathrm{percentile \ range} = 0.003 - 0.1$), indicate a marginally significant positive correlation. 
    Among the shared objects between the X-Shooter and Gaia accretion luminosity dataset, 96\% is consistent within 1$\sigma$ uncertainty. 
    We tested whether the observational errors on the Gaia-based accretion luminosities are sufficiently large to account for the intrinsic scatter observed in their relation with the X-Shooter measurements. 
    To do this, we assumed that $L_{\mathrm{acc, XShooter}}$ represents the true value. We then compared the distribution of deviations obtained when using $L_{\mathrm{acc,Gaia}}$ instead (i.e., the distance from the 1:1 line)
    with the distribution of observational uncertainties associated with $L_{\mathrm{acc,Gaia}}$. Evaluating the reduced $\chi^{2}$ and the p-value, we found $\simeq 0.63$ and $\simeq 0.98$ respectively, which indicate that the scatter between the two measurements is consistent with the observational errors in $L_{\mathrm{acc,Gaia}}$.
    The marginal correlation between the two measurements indicates that, although the Gaia-based measurements retain information on the stellar accretion, the observational uncertainties are significantly large and spectroscopic-based measurements are needed in the future to complement our analysis and accurately access the role of the external UV radiation field on stellar accretion properties. 
    Direct comparison between accretion luminosities derived from Gaia and X-Shooter spectra requires quantitative knowledge of observational biases in the Orion region, which are not well constrained. Therefore, we encoded the detection fraction by using a parametric model based on the relation with stellar luminosities, which is more appropriate than assuming a trend based on the comparison with X-Shooter data.
    We highlight that we used Gaia-based accretion luminosities in our investigation because this sample is spread across the entire Orion region and spans a wide range of FUV fluxes.  

\section{Effect of varying $M_{\mathrm{d,0}}$ and $R_{\mathrm{c},0}$ in the disc evolution model}\label{appendix:lacc_fuv_mr}
\begin{figure}
        \centering
        \includegraphics[width=0.36\textwidth]{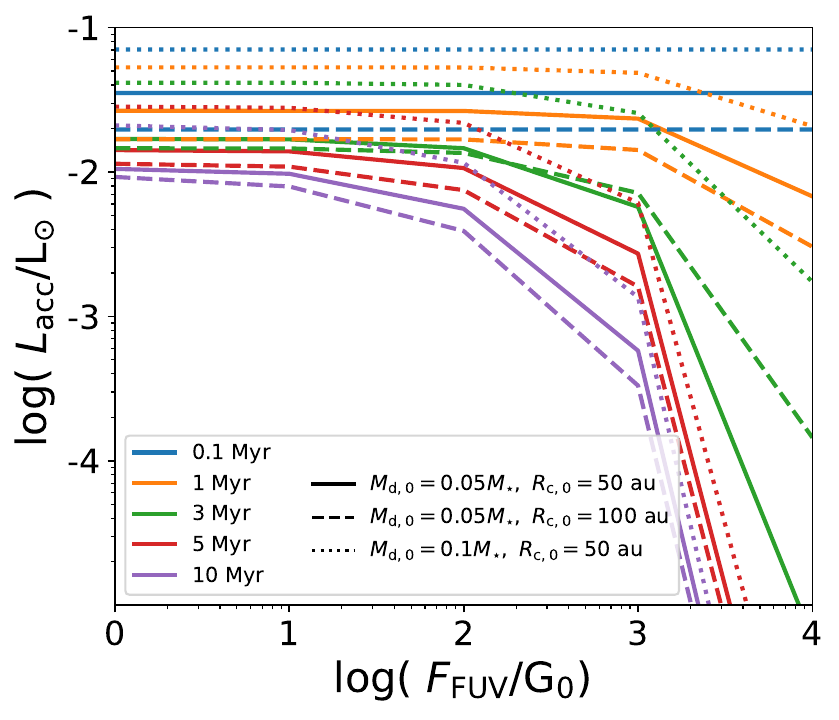}
        \caption{Evolution of $L_{\mathrm{acc}}$ across time when different FUV fluxes are included in the model, varying initial disc mass, $M_{\mathrm{d,0}}$, and initial characteristic radius, $R_{\mathrm{c},0}$. The model refers to the evolution of a disc with $M_{\star} = 0.6 \ \mathrm{M}_{\odot}$ and $a\alpha = 10^{-3}$. Different colours indicate different FUV fluxes at the outer disc boundary. Solid lines refer to $M_{\mathrm{d,0}} = 0.05 M_{\star}$ and $R_{\mathrm{c},0} = 50$ au, dashed lines corresponds to $M_{\mathrm{d,0}} = 0.05 M_{\star}$ and $R_{\mathrm{c},0} = 100$ au, and dotted lines indicates $M_{\mathrm{d,0}} = 0.1 M_{\star}$ and $R_{\mathrm{c},0} = 50$ au.}
        \label{fig:lacc_fuv_mr}
    \end{figure}

The initial conditions of the disc evolution model described in Sec. \ref{sec:method_models} include a single value for the initial disc mass, $M_{\mathrm{d,0}}=0.05M_{\star}$, and the initial characteristic radius, $R_{\mathrm{c,0}} = 50$ au. We used these representative values instead of exploring a larger parameter space because their variation do not significantly influence the time evolution of $L_{\mathrm{acc}}$ across FUV fluxes, which is the relation we aimed to explore in this work.
In Fig. \ref{fig:lacc_fuv_mr}, we show the evolution of $L_{\mathrm{acc}}$ over time in the pure viscous scenario (0 $\mathrm{G}_{0}$), and when including external photoevaporation with different levels of the FUV radiation field ($10 \ \mathrm{G}_{0}, 100 \ \mathrm{G}_{0}, 1000 \ \mathrm{G}_{0}, 10000 \ \mathrm{G}_{0}$), for a test simulated protoplanetary disc in which $M_{\star}=0.6 \ \mathrm{M}_{\odot}$ and $\alpha=10^{-3}$. Figure \ref{fig:lacc_fuv_mr} shows that the dependence of $L_{\mathrm{acc}}$ on time and FUV flux is not affected by the variation of the initial disc mass and the initial characteristic radius, instead the relation is shifted to greater or lower $L_{\mathrm{acc}}$. Based on these results, we expect that the variation of initial disc mass and extent will not drastically impact the results of our analysis. We also show that the dependence of $L_{\mathrm{acc}}$ on the FUV flux is weak for lower values of the FUV flux and in the first million years of evolution. Therefore, on a first order basis, we may expect to detect a lower $L_{\mathrm{acc}}$ for old discs exposed to different FUV fluxes. To provide a more reliable comparison with the observations of $L_{\mathbf{acc}}$ for objects in Orion, we used a population synthesis model.

\section{Inferring upper limits $L_{\mathrm{acc}}$ from the detection curve}\label{appendix:lacc_upp}
We modelled the intrinsic relation between accretion luminosity and stellar luminosity using the following procedure.
For each source $i$ we define the parameters
\begin{equation}
  x_i \equiv \log_{10}\!\left(\frac{L_{\star,i}}{L_\odot}\right), \quad \ell_{i,\mathrm{true}} \equiv \log_{10}\!\left(\frac{L_{\mathrm{acc},i}}{L_\odot}\right).
\end{equation}
The true stellar luminosities, $L_{\star}$, are not known exactly; instead we observe
$x_{i,\mathrm{obs}}$ with uncertainty $\sigma_{x,i}$. Therefore, we treated the true stellar luminosity as a latent variable, $x_{i,\mathrm{true}} \sim
\mathcal{N}\!\bigl(x_{i,\mathrm{obs}},\,\sigma_{x,i}\bigr)$,
and placed a linear model with intrinsic scatter on the true accretion luminosity,
\begin{equation}
  \mu_L(x_{i,\mathrm{true}}) = \alpha_0 + \alpha_1\bigl(x_{i,\mathrm{true}} - x_{\rm ref}\bigr), \quad 
  \ell_{i,\mathrm{true}} \sim \mathcal{N}\!\bigl(\mu_L(x_{i,\mathrm{true}}),\,\sigma_L\bigr),
\end{equation}
where $\alpha_0$, $\alpha_1$, and the intrinsic scatter $\sigma_L$ (in dex) are free parameters, and $x_{\rm ref}$ is a fixed reference luminosity (we take $x_{\rm ref}$ to be the mean of $x_{i,\mathrm{obs}}$).

The detection of accretion emission is a function of stellar luminosity. Since it is not trivial to determine the shape of this function a priori, we modelled it by employing reasonable assumptions. At the low luminosity end, the accretion luminosity may be undetected because emission is weak, and drops below some threshold sensitivity. In the intermediate regime, the detection limit may scale roughly with the stellar luminosity. At the highest stellar luminosities, the stellar spectrum may peak at shorter wavelengths associated with accretion signatures, and this may further decrease sensitivity to these accretion signatures.

We therefore encoded completeness through a parametric detection curve in accretion luminosity.
Specifically, for a given stellar luminosity $L_{\star}$ we defined the limiting accretion luminosity as
\begin{equation}
  L_{\rm lim}(L_\star)
  = L_0
    + f\,L_\star\left[1+\left(\frac{L_\star}{L_{\rm thr}}\right)^2\right],
  \label{eq:Llim_def}
\end{equation}
with $L_0$ and $L_{\rm thr}$ in units of $L_\odot$ and $f$ dimensionless. The square law of the second brackets is arbitrary, but corresponds to a steepening of the detection limit at high $L_{\star}$. The corresponding limit parameter in dex is
\begin{equation}
  \ell_{\rm lim}(L_\star)
  \equiv \log_{10}\!\left(\frac{L_{\rm lim}(L_\star)}{L_\odot}\right)
  = \log_{10} L_{\rm lim}(L_\star).
\end{equation}
Given the true accretion luminosity $\ell_{i,\mathrm{true}}$ and
$L_{\star,i}$, we assumed a ``soft'' transition between non-detections and detections. This softness is due to the fact that Gaia XP spectra quality varies between sources. The probability that a source $i$ is detected is
\begin{equation}
  z_i = \frac{\ell_{i,\mathrm{true}} - \ell_{{\rm lim},i}}{\sigma_{\rm det}}, \quad 
  p_i \equiv P(\mathrm{det}_i = 1 \mid \ell_{i,\mathrm{true}},L_{\star,i})
      = \Phi(z_i),
\end{equation}
where $\Phi$ is the standard normal cumulative distribution function
and $\sigma_{\rm det}$ (in dex) controls the width of the transition.
The observed detection flag $y_i\in\{0,1\}$ is then modelled as
\begin{equation}
  y_i \sim \mathrm{Bernoulli}(p_i).
\end{equation}
For detected sources ($y_i = 1$) we also modelled the measured accretion luminosity (in dex) as
\begin{equation}
  \ell_{i,\mathrm{obs}} \sim
  \mathcal{N}\!\bigl(\ell_{i,\mathrm{true}},\,s_{\ell,i}\bigr),
\end{equation}
where $s_{\ell,i}$ is the measurement uncertainty on $\ell_{i,\mathrm{obs}}$
(if individual errors are not available, we adopted a fixed value
$s_{\ell,i} \approx 0.05$\,dex). For non-detections ($y_i = 0$) the
numerical value of $L_{\mathrm{acc}}$ is not used in the likelihood; only
the detection flag enters via the Bernoulli term above.

We inferred the parameters $\Theta = \{\alpha_0,\alpha_1,\sigma_L,L_0,f,L_{\rm thr},\sigma_{\rm det}\}$ using the joint likelihood of detection flags
and the measured accretion luminosities for the detected sources in a Bayesian framework. We adopted weakly informative priors listed in Table \ref{tab:params}
and sampled from the posterior $P(\Theta \mid \{x_{i,\mathrm{obs}},
\sigma_{x,i}, y_i, \ell_{i,\mathrm{obs}}\})$ using the No-U-Turn Sampler
as implemented in \textsc{pymc}. Posterior summaries (medians and
credible intervals) for all parameters are reported in Table \ref{tab:params}.

In practice we adopted the posterior median values of the detection-curve parameters,
\begin{equation}
  L_0 \approx 1.6\times10^{-4}\,L_\odot,\qquad
  f \approx 4.4\times10^{-3},\qquad
  L_{\rm thr} \approx 0.42\,L_\odot,
\end{equation}
and used Eq.~\eqref{eq:Llim_def} to define an effective upper limit on the accretion luminosity of each non-detected source:
\begin{equation}
  L_{\mathrm{acc},i}^{\rm UL}
  \equiv L_{\rm lim}(L_{\star,i}; L_0,f,L_{\rm thr}),
  \qquad
  \ell_{i}^{\rm UL} = \log_{10} L_{\mathrm{acc},i}^{\rm UL}.
\end{equation}
By construction, $L_{\mathrm{acc},i}^{\rm UL}$ is the accretion luminosity
for which a source with stellar luminosity $L_{\star,i}$ has a 50\%
probability of being detected in our survey, given the jointly inferred intrinsic distribution and completeness model.

We verified on a synthetic sample that using this approach we reproduce the detection fraction, with a broadly similar distribution of detected luminosities. Therefore, for our purposes we adopt these upper limits to compare our model with observations. Refining the model for the upper limit provided by Gaia XP spectra will be topic of future work.

\begin{table}[]
    \caption{Prior and Posterior of the parameters included}
    \centering
    \begin{tabular}{c|c|c}
    \hline\hline
    Parameter & Prior & Values \\
    \hline
    $\alpha_0$ & $\mathcal{N}(-3,\,1^2)$ & $-3.58 \pm 0.0521$ \\
    $\alpha_1$ & $\mathcal{N}(2,\,1^2)$ & $1.79 \pm 0.0328$ \\
    $\sigma_L$ & $\mathrm{HalfNormal}(0.2)$ &$0.482 \pm 0.0228$ \\
    $\log_{10} (L_0/\mathrm{L}_{\odot})$ & $\mathcal{N}(-4,\,0.3^2)$& $-3.79 \pm 0.145$ \\
    $\log_{10} f$ & $ \mathcal{N}(-2,\,0.2^2)$ &$-2.35 \pm 0.0486$ \\
    $\log_{10} (L_{\rm thr}/\mathrm{L}_{\odot})$ & $\mathcal{N}\!\bigl(\log_{10} 3,\,0.15^2\bigr)$ &$-0.375 \pm 0.0343$ \\
    $\sigma_{\rm det}$ & $\mathrm{HalfNormal}(0.15)$ &$0.439 \pm 0.045$ \\
    \hline
    \end{tabular}
    \tablefoot{The parameters inferred by the Bayesian framework are listed in the left column. The central column includes the weakly informative prior used in the model, while the right column provides the posterior parameters as medians and credible intervals.}
    \label{tab:params}
\end{table}

\clearpage

\end{appendix}
\end{document}